\documentclass[journal, twocolumn]{IEEEtran}
\usepackage{amsfonts}
\usepackage{amssymb}
\usepackage{amsmath}
\usepackage{color}      
\usepackage{epsfig}
\usepackage{psfig,subfigure}

\newcommand{\calA}{\mathcal{A}}

\newcommand{\calN}{\mathcal{N}}

\def\wt{\widetilde}

\newtheorem{claim}{Claim}

\newtheorem{definition}{Definition}
\newtheorem{theorem}{Theorem}
\newtheorem{lemma}{Lemma}

\begin{document}

\title{Minimum cost mirror sites using network coding: Replication vs. coding at the source nodes}
\vspace{-0.2in}
\author{\IEEEauthorblockN{Shurui Huang, Aditya Ramamoorthy and Muriel M\'{e}dard}\\
\thanks{S. Huang and A. Ramamoorthy (\{hshurui, adityar\}@iastate.edu) are with the Dept. of Electrical \& Computer Eng., Iowa State University, Ames, IA 50011, USA. M. M\'{e}dard (medard@mit.edu) is with the Dept. of Electrical Eng. and Computer Science, Massachusetts Institute of Technology, Cambridge, MA 02139. The material in this work has appeared in part at the 2010 IEEE Information Theory Workshop, Cairo, Egypt. This work was supported in part by NSF grant CNS-0721453 and NSF grant CCF-1018148.}
}
%

\maketitle \vspace{-0.8in}
\begin{abstract}
%
%
Content distribution over networks is often
achieved by using mirror sites that hold copies of files or
portions thereof to avoid congestion and delay issues arising from
excessive demands to a single location. Accordingly, there are
distributed storage solutions that divide the file into pieces and
place copies of the pieces (replication) or coded versions of the
pieces (coding) at multiple source nodes.

We consider a network which uses network coding for multicasting
the file. There is a set of source nodes that contains either
subsets or coded versions of the pieces of the file. The cost of a
given storage solution is defined as the sum of the storage cost
and the cost of the flows required to support the multicast. Our
interest is in finding the storage capacities and flows at minimum
combined cost. We formulate the corresponding optimization
problems by using the theory of information measures. In
particular, we show that when there are two source nodes, there is
no loss in considering subset sources. For three source nodes, we
derive a tight upper bound on the cost gap between the coded and
uncoded cases. We also present algorithms for determining the
content of the source nodes.

\end{abstract}

\IEEEpeerreviewmaketitle
\begin{IEEEkeywords}
Content distribution, information measures, minimum cost, network coding.
\end{IEEEkeywords}
\vspace{-0.1in}
\section{Introduction}

Large scale content distribution over the Internet is a topic of
great interest and has been the subject of numerous studies
\cite{kirpal00}\cite{p2pperfomance}\cite{filestorage}\cite{zhanghui03}.
The dominant mode of content distribution is the client-server
model, where a given client requests a central server for the
file, which
then proceeds to service the request. 
A single server location, however is likely to be overwhelmed when
a large number of users request for a file at the same time,
because of bottleneck constraints at a storage location or other
network limitations in reaching that server location. Thus,
content, such as websites or videos for download, are often
replicated by the use of mirrors \cite{kirpal00}. Such issues are
of particular interest to Content Delivery Networks (CDNs)
\cite{CDN1}\cite{CDN2}\cite{CDN3}, which have their own, often
multi-tiered, mirroring topology. In other cases, content is
hosted by third parties, who manage complex mirroring networks and
direct requests to different locations according to the current
estimate of the Internet's congestion, sometimes termed the
weathermap, e.g., reference \cite{akamai} describes techniques for
load balancing in a network to avoid hot spots. One may consider
the usage of coding for replicating the content, e.g., through
erasure codes such as Reed-Solomon codes or fountain codes.

Peer-to-peer networks have also been proposed for content
distribution in a distributed manner
\cite{p2pperfomance}\cite{filestorage}\cite{zhanghui03}\cite{ratnasamy01}.
However, the underlying content distribution mechanism in a
peer-to-peer network is different when compared to CDNs, since
they do not use mirror sites. Instead, a given node downloads data
from available peers in a highly opportunistic fashion.
The technique of network coding has also been used for content
distribution in networks \cite{contentdist}. Under
network coding based multicast, the problem of allocating
resources such as rates and flows in the network can be solved in
polynomial time \cite{lunmincost}. Coding not only allows
guaranteed optimal performance which is at least as good as
tree-based approaches \cite{jain07}, but also does not suffer from
the complexity issues associated with
Steiner tree packings. 
Moreover, one can arrive at distributed solutions to these
problems \cite{lunmincost}\cite{zhao05}. Recently, these optimization approaches have been generalized to minimize download time \cite{wuyunnanp2p}\cite{Ho09delay}.
In these approaches, the
peers, acting as source nodes, are given. The goal of the
optimization is to reduce the download time by controlling the
amount of information transmitted at different peers. As for
multicast transmission optimization, the use of coding renders the
problem highly tractable, obviating the difficult combinatorial
issues associated with optimization in uncoded peer to peer
networks \cite{ezovski09}.

In this work, we consider the following problem. Suppose that
there is a large file, that may be subdivided into small pieces,
that needs to be transmitted to a given set of clients over a
network using network coding. The network has a designated set of
nodes (called source nodes) that have storage space. Each unit of
storage space and each unit of flow over a certain edge has a
known linear cost. We want to determine the optimal storage
capacities and flow patterns over the network such that this can
be done with minimum cost. Underlying this optimization is the
fact that source coding and network coding are not separable
\cite{adisepDSC}. Hence, there is a benefit in jointly considering
network coding for distribution and the correlation among the
sources (see \cite{liR10_chapter} for a survey). Lee et al. \cite{netcod07} and Ramamoorthy et al.
\cite{ramamoorthy09}, showed how to optimize multicast cost when
the sources are correlated. While that problem is closely related
to ours, since it considers correlated sources and optimization of
delivery using such correlated sources, it assumes a given
correlation, and no cost is associated with the storage. In this
work, we are interested in the problem of design of sources.

\begin{figure}[t]
\begin{center}
\includegraphics[width=97mm,clip=false, viewport=0 0 620 250]{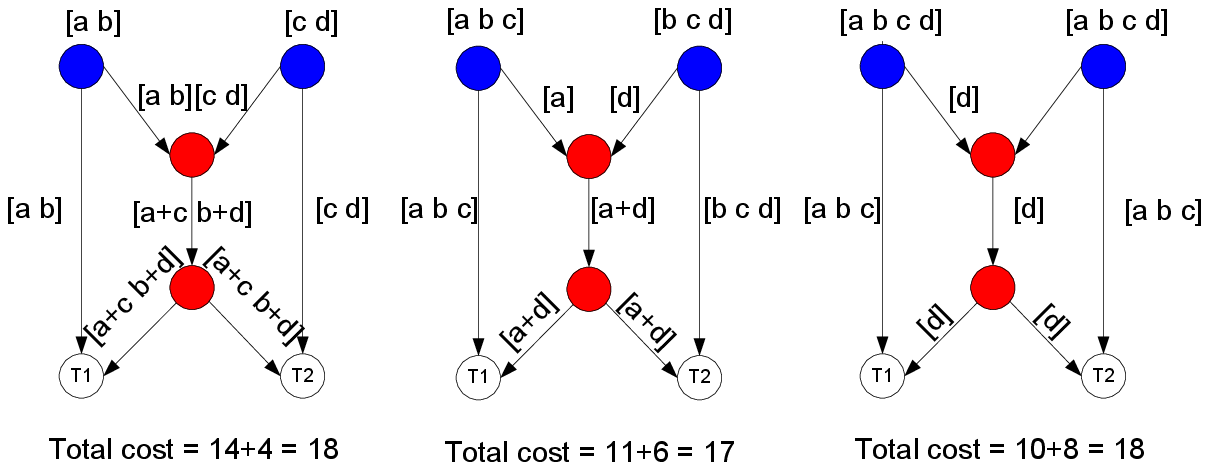}
\caption{Cost comparison of three different storage schemes when a
document $[a~b~c~d]$ needs to be transmitted to two terminals. Note that in this example, the case of partial replication has the lowest cost.}
\label{fig:intro_eg}\centering
\end{center}
\end{figure}

We distinguish the following two different cases.
\begin{itemize}
\item[(i)] \textit{Subset sources case:} Each source node only
contains an uncoded subset of the pieces of the file. \item[(ii)]
\textit{Coded sources case:} Each source node can contain
arbitrary functions of the pieces of the file.
\end{itemize}
We begin by showing by means of an example that storing
independent data at each source node is not optimal in general as
illustrated in Figure \ref{fig:intro_eg}, which is the celebrated butterfly
network. 
We consider a file represented as $(a, b, c,
d)$, where each of the four components has unit-entropy, and a
network where each edge has capacity of three bits/unit time. The cost of transmitting
at rate $x$ over edge $e$ is  $c_e(x)=x$, the cost of storage at
the sources is 1 per unit storage. As shown in the figure, the
case of partial replication when the source nodes contain
dependent information has lower cost compared to the cases when
the source nodes contain independent information or identical
information (full replication).
The case of subset sources is interesting for multiple reasons.
For example, it may be the case that a given terminal is only
interested in a part of the original file. In this case, if one
places coded pieces of the original file at the source nodes, then
the terminal may need to obtain a large number of coded pieces
before it can recover the part that it is interested in. In the
extreme case, if coding is performed across all the pieces of the
file, then the terminal will need to recover all the sources
before it can recover the part it is interested in. Note however,
that in this work we do not explicitly consider scenarios where a
given terminal requires parts of the file. From a theoretical
perspective as well, it is interesting to examine how much loss
one incurs by not allowing coding at the sources.
\subsection{Main Contributions}
%
\subsubsection{Formulation of the optimization problems by exploiting the properties of information measures (\cite{first_info})} We provide a precise formulation of the different optimization problems by leveraging the properties of the information measure (I-measure) introduced in \cite{first_info}. This allows to provide a succinct formulation of the cost gap between the two cases and allows us to recover tight results in certain cases.
\subsubsection{Cost comparison between subset sources case and coded sources case}
The usage of the properties of information measure allows us to
conclude that when there are two source nodes, there is no loss in
considering subset sources. Furthermore, in the case of three
source nodes, we derive an upper bound on the cost between the two
cases that is shown to be tight. Finally, we propose a greedy
algorithm to determine the cost gap for a given instance.

This paper is organized as follows. In Section
\ref{sec:background}, we present background and related work.
Section \ref{sec:preliminaries} outlines basic results that allow
us to apply the theory of I-measures to our problem. We formulate
the precise problems under consideration in Section
\ref{sec:problem}. The cost gap between the subset case and the
coded case is discussed in Section \ref{sec:cost}, and the
simulation results are presented in Section \ref{sec:results}.
Section \ref{sec:conclu} concludes the paper. \vspace{-0.2in}
\section{Background and related work}
\label{sec:background}
\subsection{Minimum cost multicast with multiple sources problem}
Several schemes have been proposed for content distribution over
networks as discussed previously
(\cite{kirpal00}\cite{filestorage}\cite{zhanghui03}\cite{ratnasamy01}\cite{contentdist}).
In this section we briefly overview past work that is most closely
related to the problem that we are considering.

%

Network coding has been used in the area of large scale content distribution for different purposes.
Several design principles for peer to peer streaming system with network coding in realistic settings are introduced in \cite{baochunli08}. Reference \cite{contentdist} proposed a content distribution scheme using network coding in a dynamic environment where nodes cooperate. 
A random linear coding based storage system (which is motivated by random network coding) was considered in \cite{netcod2005_accek} and shown to be more efficient than uncoded random storage system. However, their notion of efficiency is different than the total flow and storage cost considered in our work. 
The work of \cite{lunmincost}, proposed linear programming
formulations for minimum cost flow allocation network coding based
multicast. Lee et al. \cite{netcod07} constructed minimum cost
subgraphs for the multicast of two correlated sources. They also
proposed the problem of optimizing the correlation structure of
sources and their placement. However, a solution was not presented
there. Efficient algorithms for jointly allocating flows and rates
were proposed for the multicast of a large number of correlated
sources by Ramamoorthy \cite{ramamoorthy09} (see \cite{ramamoorthyRS09} for a formulation where the terminals exhibit selfish behavior). The work of Jiang
\cite{jointopt}, considered a formulation that is similar to ours.
It shows that under network coding, the problem of minimizing the
joint transmission and storage cost can be formulated as a linear
program. Furthermore, it considers a special class of networks
called generalized tree networks and shows that there is no
difference in the cost whether one considers subset sources or
coded sources. This conclusion is consistent with the fact that
network coding is not useful in tree settings. In contrast, in
this work we consider general networks, i.e., we do not assume any
special structure of the network. We note that in more recent work \cite{NCDSS}, network coding based distributed storage mechanisms and associated research issues have been outlined.

The work of Bhattad et al. \cite{minimal_nc} proposed an optimization problem formulation for cost minimization when some nodes are only allowed routing and forwarding instead of network coding. Our work on subset sources can perhaps be considered as an instance of this problem, by introducing a virtual super node and only allowing routing/forwarding on it. However, since we consider a specific instance of this general problem, as we allow coding at all nodes except the virtual super node, our problem formulation is much simpler than \cite{minimal_nc} and allows us to compare the cost of subset sources vs. coded sources.  In \cite{minimal_nc}, the complexity grows as the product of the number of edges and a factor that is exponential in the number of terminals. In our method, the number of constraints only grows linearly with the number of receivers. However, there is a set of constraints that is exponential in the number of source nodes. 
For most networks, we expect our formulation to be more efficient.
In addition, we recover stronger results in the case when there are only two or three source nodes.
Our solution approach uses the concept of information measures \cite{first_info}, that has also been used in \cite{NormalizedEV} recently in other contexts.

\subsection{Set theory and information theory}
\label{sec:pre}
In this section, we introduce a few basic concepts and useful theorems that relate to set theory and information theory. More details can be found in \cite{first_info}.
\begin{definition}
The field $\mathcal F_n$ generated by sets $\widetilde{X}_1,\widetilde{X}_2,\cdots,\widetilde{X}_n$ is the collection of sets which can be obtained by any sequence of usual set operations on $\widetilde{X}_1,\widetilde{X}_2,\cdots,\widetilde{X}_n$.
\end{definition}
\begin{definition}
The atoms of $\mathcal{F}_n$ are sets of the form $\cap_{i=1}^nY_i$, where $Y_i$ is either $\widetilde{X}_i$ or $\widetilde{X}_i^c$.
\end{definition}

\begin{definition}
A real function $\mu$ defined on $\mathcal F_n$ is called a signed measure if it is set-additive, i.e., for disjoint sets A and B in $\mathcal F_n$, $\mu(A\cup B)=\mu(A)+\mu(B)$. 
\end{definition}

We use $\mathcal F_n$ to denote the field generated by $\widetilde{X}_1,\widetilde{X}_2,\cdots,\widetilde{X}_n$. Define the universal set $\Omega$ to be the union of the sets $\widetilde{X}_1,\widetilde{X}_2,\cdots,\widetilde{X}_n$, i.e., $\Omega=\cup_{i=1}^n\widetilde{X}_i$. The set $A_0=\cap_{i=1}^n\widetilde{X}_i^c$ whose measure is $\mu(\cap_{i=1}^n\widetilde{X}_i^c)=\mu(\emptyset)=0$, is called the empty atom of $\mathcal F_n$. Let $\mathcal A$ be the set of nonempty atoms of $\mathcal F_n$ ($|\mathcal A|=2^n-1$). It can be shown that any set in $\mathcal F_n$ can be uniquely defined as the union of some atoms. A signed measure $\mu$ on $\mathcal F_n$ is completely specified by the values of the $\mu$ on the nonempty atoms of $\mathcal F_n$.

Consider a field $\mathcal F_n$ generated by $n$ sets $\widetilde{X}_1,\widetilde{X}_2,\cdots,\widetilde{X}_n$. Let $\mathcal{N}_S=\{1,2,\cdots,n\}$ and $\widetilde{X}_V$ denote $\cup_{i\in V}\widetilde{X}_i$ for any nonempty subset $V$ of $\mathcal N_S$.
Define $
\mathcal B=\{\widetilde{X}_V:~V\text{ is a nonempty subset of }\mathcal N_S\}
$ According to the proof of Theorem 3.6 in \cite{first_info}, there is a unique linear relationship between $\mu(A)$ for $A\in\mathcal A$ and $\mu(B)$ for $B\in\mathcal B$.
Since $\mathcal F_n$ can be completely specified by $\mu(A)$, $\mathcal F_n$ can also be completely specified by $\mu(B)$.

For $n$ random variables $X_1,X_2,\cdots,X_n$, let $\widetilde{X}_i$ be a set corresponding to $X_i$. Let $X_V=(X_i,i\in V)$, where $V$ is some nonempty subset of $\mathcal N_s$. We define the signed measure by $\mu^*(\widetilde X_V)=H(X_V)$, for all nonempty subset $V$ of $\mathcal N_S$. Then
$\mu^*$ is the unique signed measure on $\mathcal F_n$ which is consistent with all of Shannon's information measures (Theorem 3.9 in \cite{first_info}).


\section{Preliminaries}
\label{sec:preliminaries}
In this section we develop some key results, that will be used throughout the paper. In particular, we shall deal extensively with the I-measure introduced in \cite{first_info}. We refer the reader to \cite{first_info} for the required background in this area. First we note that it is well known that atom measures can be negative for general probability distributions \cite{first_info}, e.g., three random variables $X_1$, $X_2$ and $X_3$, where $X_1$ and $X_2$ are independent, $P(X_i=1)=P(X_i=0)=1/2$, $i=1,2$. $X_3=(X_1+X_2)\mod2$, then $\mu(\widetilde{X}_1\cap\widetilde{X}_2\cap\widetilde{X}_3)=-1$. Next we argue that in order to make each source node only contain a subset of the pieces of the file, the measure of the atoms in the fields generated by the sources should be non-negative. This is stated as a theorem below.

Let $\mathcal N_S=\{1,2,\cdots,n\}$. Consider $n$ random variables $X_1,X_2,\cdots, X_n$ and their corresponding sets $\widetilde{X}_1,\widetilde{X}_2,\cdots,\widetilde{X}_n$. Let $\wt{X}_V = \cup_{i\in V}\wt{X}_i$ and $X_V=(X_i,i\in V)$, $V\subseteq\calN_S$. We denote the set of nonempty atoms of $\mathcal F_n$ by $\mathcal A$, where $\mathcal F_n$ is the field generated by the sets $\widetilde{X}_1,\widetilde{X}_2,\cdots,\widetilde{X}_n$. Construct the signed measure $\mu^*(\wt{X}_V)=H(X_V)$, for all nonempty subset $V$ of $\calN_S$.


\begin{theorem}
\label{th:sb}
(1) Suppose that there exists a set of $2^n - 1$ nonnegative values, one corresponding to each atom of $\mathcal{F}_n$, i.e, $\alpha(A) \geq 0, \forall A \in \mathcal{A}$. Then, we can define a set of independent random variables, $W_A, A \in \calA$ and construct random variables $X_j = (W_A : A \in \calA, A \subset \wt{X}_j$), such that the measures of the nonempty atoms of the field generated by $\widetilde{X}_1,\widetilde{X}_2,\cdots,\widetilde{X}_n$ correspond to the values of $\alpha$, i.e., $\mu^*(A) = \alpha(A), \forall A \in \calA$. \\ (2) Conversely, let $Z_i, i \in \{1, \dots, m\}$ be a collection of independent random variables. Suppose that a set of random variables $X_i, i = 1, \dots, n$ is such that $X_i = Z_{V_i}$, where  $V_i \subseteq \{1, \dots, m\}$. Then the set of atoms of the field generated by $\widetilde{X}_1,\widetilde{X}_2,\cdots,\widetilde{X}_n$, have non-negative measures.
\end{theorem}
\emph{Proof:} See Appendix.
\endproof

\section{Problem Formulation}
\label{sec:problem}

We now present the precise problem formulations for the subset sources case and the coded sources case.
Suppose that we are given a directed graph $G=(V,E,C)$ that represents the network, $V$ denotes the set of vertices, $E$ denotes the set of edges, and $C_{ij}$ denotes the capacity of edge $(i,j)\in E$. There is a set of source nodes $S \subset V$ (numbered $1, \dots, n$) and terminal nodes $T \subset V$, such that $|T|=m$. We assume that the original source, that has a certain entropy, can be represented as the collection of equal entropy independent sources $\{OS_j\}_{j=1}^Q$, where $Q$ is a sufficiently large integer. Note that this implies that $H(OS_j)$ can be fractional. Let $X_i$ represent the source at the $i^{th}$ source node. For instance in the case of subset sources, this represents a subset of $\{OS_j\}_{j=1}^Q$ that are available at the $i^{th}$ node. Suppose that each edge $(i,j)$ incurs a linear cost $f_{ij} z_{ij}$ for a flow of value $z_{ij}$ over it, and each source incurs a linear cost $d_i H(X_i)$ for the information $X_i$ stored.
\subsection{Subset Sources Case}
\subsubsection{Basic formulation}
In this case each source $X_i, i = 1, \dots, n$ is constrained to be a subset of the pieces of the original source. We leverage Theorem \ref{th:sb} from the previous section that tells us that in this case that $\mu^*(A) \geq 0$ for all $A \in \calA$. In the discussion below, we will pose this problem as one of recovering the measures of the $2^n - 1$ atoms. Note that this will in general result in fractional values. However, the solution can be interpreted appropriately because of the assumptions on the original source. This point is also discussed in Section \ref{sec:soln_explanation}.



\begin{figure}[t]
\begin{center}
\includegraphics[width=80mm,clip=false, viewport=10 0 280 260]{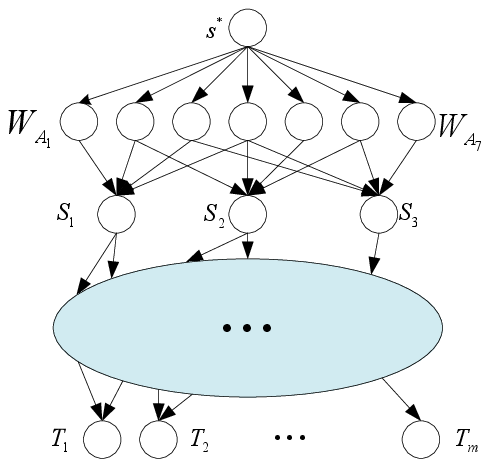}
\caption{Modified graph for the first formulation when there are three sources.}
\label{fig:at_stru}\centering
\end{center}
\end{figure}

We construct an augmented graph $G^*_1=(V^*_1,E^*_1, C^*_1)$ as follows (see Figure \ref{fig:at_stru}). Append a virtual super node $s^*$ and $2^n-1$ virtual nodes corresponding to the atom sources $W_A, \forall A \in \calA$ and connect $s^*$ to each $W_A$ source node. 
The node for $W_A$ is connected to a source node $i \in S$ if $A \subset \wt{X}_i$. The capacities of the new (virtual) edges are set to infinity. The cost of the edge $(s^*, W_A)$ is set to $\sum_{\{i \in S: A \subset \wt{X}_i\}} d_i$. The costs of the edges $(W_A,S_i),A \subset \wt{X}_i$ are set to zero.


If each terminal can recover all the atom sources, $W_A, \forall A \in \calA$, then it can in turn recover the original source. The information that needs to be stored at the source node $i \in S$, is equal to the sum of flows from $s^*$ to $W_A, \forall A\subset \wt{X}_i$.  Let $x_{ij}^{(t)}$, $t\in T$ represent the flow variable over $G^*_1$ corresponding to the terminal $t$ along edge $(i,j)$ and let $z_{ij}$ represent $\max_{t \in T} x_{ij}^{(t)},\forall (i,j)\in E$. The corresponding optimization problem is defined as ATOM-SUBSET-MIN-COST. 

\begin{figure}[t]
\begin{center}
\includegraphics[width=80mm,clip=false, viewport=0 0 280 270]{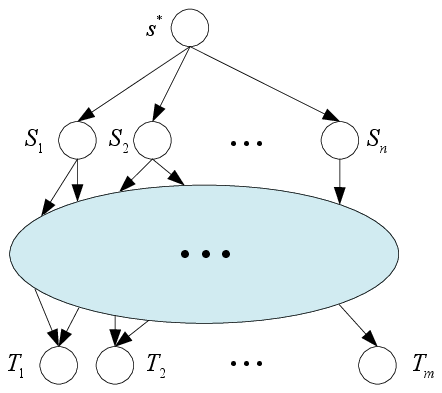}
\caption{Modified graph for the second formulation.}
\label{fig:sour_stru}\centering
\end{center}
\end{figure}


minimize $\sum_{(i,j)\in E}f_{ij}z_{ij}+\sum_{A\in \calA}(\sum_{\{i \in S: A \subset \wt{X}_i\}} d_i)\mu^*(A)$

subject to
\begin{align}
&0\leq x^{(t)}_{ij}\leq z_{ij}\leq c_{ij,1}^*,\forall (i,j)\in E^*_1, t\in T\notag\\
&\sum_{\{j|(i,j)\in E^*_1\}}x_{ij}^{(t)}-\sum_{\{j|(j,i)\in E^*_1\}}x_{ji}^{(t)}=\sigma_i^{(t)},\forall i\in V^*_1,~t\in T\notag\\
\label{eq:atom}
&x^{(t)}_{s^*W_A}=\mu^*(A), t\in T, A\in \calA\\
\label{eq:subset1}
&\mu^*(A)\geq0, \forall A\in\calA\\
\label{eq:joint}
&H(X_1,X_2\cdots,X_n)=\sum_{A:A\in\calA}\mu^*(A)
\end{align}
where
\begin{equation}
\label{eq:sigma}
\sigma_i^{(t)}=\left\{ \begin{array}{ll}
H(X_1,\cdots, X_n)&\mbox {if $i=s^*$ }\\
-H(X_1,\cdots, X_n)& \mbox {if $i=t$ }\\
0& \mbox {otherwise.}
\end{array} \right.
\end{equation}
This is basically the formulation of the minimum cost multicast problem \cite{lunmincost} with a virtual super-source of entropy $H(X_1, \dots, X_n)$, with the added constraint that the flow on the edge from $s^*$ to node $W_A$ for each terminal, $x^{(t)}_{s^* W_A}$ is at least $\mu^*(A)$. We also have a constraint that $\sum_{A\in\calA}\mu^*(A) = H(X_1,X_2,\cdots,X_n)$, that in turns yields the constraint that $x^{(t)}_{s^* W_A} = \mu^*(A)$. Also, note that the measure of each atom, $\mu^*(A)$ is non-negative. This enforces the subset constraints. Because from the non-negative measures of the atoms, we are able to construct random variables, which indicates the atom measures satisfy both Shannon type inequalities and non-Shannon type inequalities. Hence, the non-negative atom measures ensure that the corresponding entropic vectors are in entropy region.

In general, the proposed LP formulation has a number of constraints that is exponential in the number of source nodes, since there are $2^n - 1$ atoms. However, when the number of source nodes is small, this formulation can be solved using regular LP solvers. We emphasize, though, that the formulation of this problem in terms of the atoms of the distribution of the sources provides us with a mechanism for reasoning about the case of subset constraints, under network coding. We are unaware of previous work that proposes a formulation of this problem. 

In order to provide bounds on the gap between the optimal costs of the subset sources case and the coded sources case, we now present an alternate formulation of this optimization, that is more amenable to gap analysis. Note however, that this alternate formulation has more constraints than the one presented above.
\subsubsection{Another formulation}
In the first formulation, the terminals first recover the atom sources, and then the original source. In this alternate formulation, we pose the problem as one of first recovering all the sources, $X_i$, $i \in S$ at each terminal and then the original source. Note that since these sources are correlated, this formulation is equivalent to the Slepian-Wolf problem over a network \cite{ramamoorthy09}. We shall first give the problem formulation and then prove that the two formulations have the same optimums.

We construct another augmented graph $G^*_2=(V^*_2,E^*_2,C^*_2)$ (see Figure \ref{fig:sour_stru}) using the basic network graph $G=(V,E,C)$. We append a virtual super node $s^*$ to $G$, and connect $s^*$ and each source node $i$ with virtual edges, such that its capacity is infinity and its cost is $d_i$. 

As before, let $x_{ij}^{(t)}$, $t\in T$ represent the flow variable over $G^*_2$ corresponding to the terminal $t$ along edge $(i,j)$ and let $z_{ij}$ represent $\max_{t \in T} x_{ij}^{(t)},\forall (i,j)\in E$.
We introduce variable $R_i^{(t)},t\in T$ that represents the rate from source $i$ to terminal $t$, $i=1,\cdots,n$. Thus $R^{(t)}=(R^{(t)}_1,R^{(t)}_2,\cdots,R^{(t)}_n)$ represents the rate vector for terminal $t$. In order for $t$ to recover the sources \cite{Tracy06}, the rate vector $R^{(t)}$ needs to lie within the Slepian-Wolf region of the sources
\vspace{-0.1in}
\begin{equation*}
\mathcal R_{\mathcal{SW}}=\{(R_1,\cdots,R_n):\forall U\subseteq S,
\sum_{i\in U}R_i\geq H(X_U|X_{S\setminus U})\}.
\end{equation*}
Moreover, the rates also need to be in the capacity region such that the network has enough capacity to support them for each terminal.
As before we enforce the subset constraint $\mu^*(A) \geq 0, \forall A \in \calA$. The optimization problem is defined as SUBSET-MIN-COST. 
%

minimize $\sum_{(i,j)\in E}f_{ij}z_{ij}+\sum_{A\in \calA}(\sum_{\{i \in S: A \subset \wt{X}_i\}} d_i)\mu^*(A)$

subject to
\begin{align}
\label{eq:che}
&0\leq x^{(t)}_{ij}\leq z_{ij}\leq c_{ij,2}^*,~(i,j)\in E^*_2, t\in T\\
&\sum_{\{j|(i,j)\in E^*_2\}}x_{ij}^{(t)}-\sum_{\{j|(j,i)\in E^*_2\}}x_{ji}^{(t)}=\sigma_i^{(t)},~i\in V^*_2,~t\in T\nonumber\\
\label{eq:mincut}
&x^{(t)}_{s^*i}\geq R_i^{(t)},\forall i\in S, t\in T\\
\label{eq:sw}
&R^{(t)}\in \mathcal R_{\mathcal{SW}}, \forall t\in T\\
\label{eq:subset}
&\mu^*(A)\geq0,\forall A\in\calA\\
&z_{s^*i}=H(X_i)=\sum_{A:A\in\calA,A\subset \wt{X}_i}\mu^*(A),\forall i\in S\\
\label{eq:marginal}
&H(X_1,X_2,\cdots ,X_n)=\sum_{A\in\calA}\mu^*(A)\\
&H(X_U|X_{S\setminus U})=\sum_{A:A\in\calA,A\nsubseteq \wt{X}_{S\setminus U}}\mu^*(A), \forall U\subseteq S
\end{align}
where $\sigma_i^{t}$ is defined in (\ref{eq:sigma}).


Now we prove the two formulations will get the same optimal values. The basic idea is as follows. Note that the objective functions for both the formulations are exactly the same. We shall first consider the optimal solution for the first formulation and construct a solution for the second formulation so that we can conclude that $f_{opt1} \geq f_{opt2}$. In a similar manner we will obtain the reverse inequality, which will establish equality of the two optimal values.

Suppose that we are given the optimal set of flows $x_{ij,1}^{(t)},z_{ij,1}, t\in T, (i,j)\in E^*_1$ and the optimal atom values $\mu^*(A)_1$ for the first formulation, with an objective of value $f_{opt1}$.
\begin{claim}
\label{claim:1eq2}
In $G^*_2$, for the flows $x_{ij,2}^{(t)}$, $z_{ij,2}$, and the atoms $\mu^*(A)_2$, assign
\vspace{-0.1in}
$$x_{ij,2}^{(t)}=x_{ij,1}^{(t)},~ z_{ij,2}=z_{ij,1}, \forall (i,j)\in G$$
\begin{equation*}
\begin{split}
R_{i,2}^{(t)}=x_{s^*i,2}^{(t)}&=\sum_{A:A\in\calA,A\subset\wt{X}_i}x^{(t)}_{W_Ai,1}, ~z_{s^*i,2}\\
&=\sum_{A:A\in\calA,,A\subset\wt{X}_i}\mu^*(A)_1,\forall i\in S,t\in T
\end{split}
\end{equation*}
$$\mu^*(A)_2=\mu^*(A)_1,\forall A\in \calA.$$
Then $R_{i,2}^{(t)}$, $x_{ij,2}^{(t)}$, $z_{ij,2}$, and the atoms $\mu^*(A)_2$ are a feasible solution for the second formulation.
\end{claim}
\emph{Proof.}
Flow balance for source node $i\in S$ in the first formulation implies that $\sum_{A:A\subset\wt{X}_i}x^{(t)}_{W_Ai,1}=\sum_{j:(i,j)\in E_1^*}x^{(t)}_{ij,1}$, $\forall i\in S$. Therefore flow balance for source node $i$ in the second formulation can be seen as follows:  $x_{s^*i,2}^{(t)}=\sum_{A:A\in\calA,A\subset\wt{X}_i}x^{(t)}_{W_Ai,1}=\sum_{j:(i,j)\in E_1^*}x^{(t)}_{ij,1}=\sum_{j:(i,j)\in E_2^*}x^{(t)}_{ij,2},\forall i\in S$. Flow balance at the internal nodes is trivially satisfied. We only need to check constraints (\ref{eq:mincut}) and (\ref{eq:sw}).

In the equations below, we use $A \in \mathcal A$ (i.e., $A$ is an atom) as a summation index at various terms. However, for notational simplicity, we do not explicitly include the qualifier, $A\in\mathcal A$ below. Also in the equations, we have the convention that if there is no edge between nodes $W_A$ and $i$ in $G^*_1$, the flow $x^{(t)}_{W_Ai,1}$ is zero. For any $U\subseteq S$, we have
\begin{equation}
\begin{split}
&\sum_{i\in U}x_{s^*i,2}^{(t)}=\sum_{i\in U}\sum_{A:A\subset\wt{X}_i}x^{(t)}_{W_Ai,1}\stackrel{_{(a)}}{=}\sum_{i\in U}\sum_{A:A\subset\wt{X}_U}x^{(t)}_{W_Ai,1}\\
&=\sum_{i\in U}\sum_{A:A\nsubseteq \wt{X}_{S\setminus U},A\subset\wt{X}_U}x^{(t)}_{W_Ai,1}+\sum_{i\in U}\sum_{A:A\subseteq \wt{X}_{S\setminus U},A\subset\wt{X}_U}x^{(t)}_{W_Ai,1}\\
&\geq\sum_{i\in U}\sum_{A:A\nsubseteq \wt{X}_{S\setminus U},A\subset\wt{X}_U}x^{(t)}_{W_Ai,1}=\sum_{A:A\nsubseteq \wt{X}_{S\setminus U},A\subset\wt{X}_U}\sum_{i\in U}x^{(t)}_{W_Ai,1}\\
&\stackrel{_{(b)}}{=}\sum_{A:A\nsubseteq \wt{X}_{S\setminus U},A\subset\wt{X}_U}x^{(t)}_{s^*W_A,1}\stackrel{_{(c)}}{=}\sum_{A:A\nsubseteq \wt{X}_{S\setminus U},A\subset\wt{X}_U}\mu^*(A)_1\\
&=\sum_{A:A\nsubseteq \wt{X}_{S\setminus U},A\subset\wt{X}_U}\mu^*(A)_2=H(X_U|X_{S\setminus U})
\end{split}
\end{equation}
where $H(X_U|X_{S\setminus U})$ is the conditional entropy of the second formulation. $(a)$ is due to the convention we defined above.  $(b)$ is from the flow balance at the atom node and the convention we defined above. $(c)$ comes from the constraint (\ref{eq:atom}) in the first formulation. Therefore, constraints (\ref{eq:mincut}) and (\ref{eq:sw}) are satisfied and this assignment is feasible for the second formulation with a cost equal to $f_{opt1}$.
\endproof
\noindent We conclude that the optimal solution for the second formulation $f_{opt2}$ will have $f_{opt2}\leq f_{opt1}$.

Next we show the inequality in the reverse direction. Suppose that we are given the optimal set of flows $x_{ij,2}^{(t)},z_{ij,2},t\in T, (i,j)\in E^*_2$ and the atom values $\mu^*(A)_2$ in the second formulation. Further assume that the optimal objective function is $f_{opt2}$.
\begin{claim}
\label{claim:2eq1}
In $G^*_1$, assign
$$x_{ij,1}^{(t)}=x_{ij,2}^{(t)},z_{ij,1}=z_{ij,2}, \forall (i,j)\in G$$
$$z_{s^*W_A,1}=x_{s^*W_A,1}^{(t)}=\mu^*(A)_1=\mu^*(A)_2,\forall A\in\calA.$$
Furthermore, there exist flow variables $x_{W_Ai,1}^{(t)}$ and $z_{W_Ai,1}$ over the edge $(W_A,i)\in V^*_1$, $\forall A\in\calA$, such that together with the assignment, they form a feasible solution for the first formulation.
\end{claim}
\begin{figure}[htbp]
\begin{center}
\includegraphics[width=80mm,clip=false, viewport=0 0 280 160]{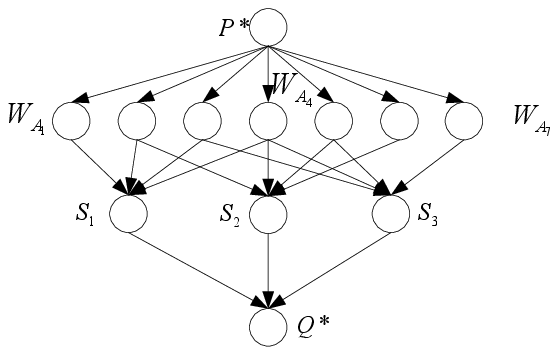}
\caption{An example of the graph constructed for the proof of Claim \ref{claim:2eq1}, where there are three sources.} \label{fig:source-atom}\centering
\end{center}
\end{figure}

\emph{Proof.} It is clear that the assignments for $x_{ij,1}^{(t)}$ and $z_{ij,1}$ for $(i,j) \in G$ satisfy the required flow balance constraints. We need to demonstrate the existence of flow variables $x_{W_Ai,1}^{(t)}$ and $z_{W_Ai,1}$ over the edge $(W_A,i)\in V^*_1$, $\forall A\in\calA$, such that they satisfy the flow balance constraints. 

Towards this end it is convenient to construct an auxiliary graph as follows. There is a source node $P^*$ connected to the atoms $W_A$'s, $A\in\calA$, a terminal $Q^*$ connected to the sources nodes, $i\in S$. There is an edge connecting $W_A$ and $i$ if $A\subset\wt{X}_i$. An example is shown in Figure \ref{fig:source-atom}. 
The capacity for edge $(P^*,W_A)$ is $x_{s^*W_A,1}^{(t)}$, the capacity for edge $(i,Q^*)$ is $x_{s^*i,2}^{(t)}$, and the capacity for edge $(W_A,i)$ is infinity. Note that $\sum_{A \in \calA} x_{s^*W_A,1}^{(t)} = \sum_{i \in S} x_{s^*i,2}^{(t)} = H(X_1, X_2, \cdots, X_n)$. Therefore, if we can show that the maximum flow in this auxiliary graph between $P^*$ and $Q^*$ is $H(X_1, X_2, \cdots, X_n)$, this would imply the existence of flow variables on the edges between the atom nodes and the source nodes that satisfy the required flow balance conditions.


To show this we use the max-flow min-cut theorem \cite{kleinburgT} and instead show that the minimum value over all cuts separating $P^*$ and $Q^*$ is $H(X_1, X_2, \cdots, X_n)$. 

First, notice that there is a cut with value $H(X_1,X_2,\cdots,X_n)$. This cut can be simply the node $P^*$, since the sum of the capacities of its outgoing edges is $H(X_1,X_2,\cdots,X_n)$. 
Next, if an atom node $W_A$ belongs to the cut that contains $P^*$, then we must have all source nodes $i\in S$ such that $A\subset\wt{X}_i$ also belonging to the cut. To see this, note that otherwise there is at least one edge crossing the cut whose capacity is infinity, i.e., the cut cannot be the minimum cut.

Let $S' \subseteq S$. Based on this argument it suffices to consider cuts that contain, $P^*$, the set of nodes $S \setminus S'$ and the set of all atoms $W_A$ such that $A\nsubseteq\wt{X}_{S'}$. The value of this cut is at least
\begin{equation*}
\begin{split}
&\sum_{A:A\in\calA,A\subseteq\wt{X}_{S'}}x_{s^*W_A,1}^{(t)}+\sum_{i\in S\setminus S'}x_{s^*i,2}^{(t)}\\
&=H(X_1,\cdots,X_n)-\sum_{A:A\in\calA,A\nsubseteq\wt{X}_{S'}}x_{s^*W_A,1}^{(t)}+\sum_{i\in S\setminus S'}x_{s^*i,2}^{(t)}.
\end{split}
\end{equation*}
\indent By constraints (\ref{eq:mincut}), (\ref{eq:sw}) and the given assignment, we have $\sum_{A:A\in\calA,A\nsubseteq\wt{X}_{S'}}x_{s^*W_A,1}^{(t)}=H(X_{S\setminus S'}|X_{S'}) \leq \sum_{i\in S\setminus S'}x_{s^*i,2}^{(t)}$. This implies that the value of any cut of this form at least $H(X_1,X_2,\cdots,X_n)$. Therefore we can conclude that the minimum cut over all cuts separating $P^*$ and $Q^*$ is exactly $H(X_1,X_2,\cdots,X_n)$, 
i.e., our assignment is a valid solution.
\endproof
Using Claims \ref{claim:1eq2} and \ref{claim:2eq1}, we conclude that $f_{opt1}=f_{opt2}$.

As mentioned earlier, the second formulation will be useful when we compute the cost gap between the coded and subset cases, we will use the graph $G^*=G^*_2$ in the rest of the paper.
\subsection{Solution explanation and construction}
\label{sec:soln_explanation}

Assume that we solve the above problem and obtain the values of all the atoms $\mu^*(A), A\in\calA$.  These will in general be fractional. We now outline the algorithm that decides the content of each source node. We use the assumption that the original source can be represented as a collection of independent equal-entropy random variables $\{OS_i\}_{i=1}^Q$, for large enough $Q$ at this point. Suppose that $H(OS_1) = \beta$. In turn, we can conclude that there exist integers $\alpha_A, \forall A \in \calA$, such that $ \alpha_A \times \beta = \mu^*(A), \forall A \in \calA$ and that $\sum_{A \in \calA} \alpha_A = Q$. Consider an ordering of the atoms, denoted as $A_1, A_2, \cdots, A_{2^n - 1}$. The atom sources can then be assigned as follows: For each $A_i$, assign $W_{A_i} = (OS_{\sum_{j < i} \alpha_{A_{j}} + 1}, OS_{\sum_{j < i} \alpha_{A_{j}} + 2}, \dots, OS_{\sum_{j \leq i} \alpha_{A_{j}}})$. It is clear that the resultant atom sources are independent and that $H(W_A) = \mu^*(A), \forall A \in \calA$. Now set $X_i=(W_A:A\subset \wt{X}_i)$, to obtain the sources at each node.

The assumption on the original source is essentially equivalent to saying that a large file can be subdivided into arbitrarily small pieces. To see this assume that each edge in the network has a capacity of 1000 bits/sec. At this time-scale, suppose that we treat each edge as unit-capacity. If the smallest unit of a file is a single bit, then we can consider it to be consisting of sources of individual entropy equal to $10^{-3}$. 
%
\subsection{Coded source network}
Given the same network, if we allow coded information to be stored at the sources, using the augmented graph $G^*$ by the second problem formulation, the storage at the sources can be viewed as the transmission along the edges connecting the virtual source and real sources. Then the problem becomes the standard minimum cost multicast with network coding problem (CODED-MIN-COST) \cite{lunmincost} where the variables are only the flows $z_{ij}$ and $x_{ij}^{(t)}$.

minimize $\sum_{(i,j)\in E}f_{ij}z_{ij}+\sum_{i\in S}d_iz_{s^*i}$

subject to 
$$0\leq x^{(t)}_{ij}\leq z_{ij}\leq c_{ij}^*,~(i,j)\in
E^*, t\in T$$
$$
\sum_{\{j|(i,j)\in E^*\}}x_{ij}^{(t)}-\sum_{\{j|(j,i)\in E^*\}}x_{ji}^{(t)}=\sigma_i^{(t)},~i\in V^*,~t\in T
$$

where $\sigma_i^{t}$ is defined in (\ref{eq:sigma}).
Assume we have the solution for CODED-MIN-COST, we can use the random coding scheme introduced by \cite{Tracy06} or other deterministic coding schemes \cite{Jaggi05} to reconstruct the sources and the information flow of each edge. 

\section{Cost comparison between the coded case and subset case}
\label{sec:cost} For given instances of the problem, we can
certainly compute the cost gap by solving the corresponding
optimization problems SUBSET-MIN-COST and CODED-MIN-COST. In this section we formulate an alternate version of CODED-MIN-COST where we also seek to obtain the values of the atom measures of the sources (as we did for SUBSET-MIN-COST). In principle, this requires us to ensure that the atom measures to satisfy
the information inequalities \cite{first_info} that consist of
Shannon type inequalities and non-Shannon type inequalities when
$n\geq4$. In reference \cite{infinite_matus}, it was shown that
there are infinitely many non-Shannon type inequalities when
$n\geq 4$. Hence, it is impossible to list all the information
inequalities when the source number exceeds 4. Moreover, since the entropic region is not polyhedral, the problem is no longer an LP. In our optimization we only enforce the Shannon inequalities and remove the non-negativity constraint on the atom measures. In general, these atom measures may not correspond to the distribution of an actual coded solution. However, as explained below, starting with an output of our LP, we find a feasible instance for the SUBSET-MIN-COST problem and then arrive at an upper bound on the gap. 

In the general case, of $n$ sources, even this optimization has constraints that are exponential in $n$. However, this formulation still has advantages. In particular, we are able to provide a greedy algorithm to find near-optimal solutions for it. Moreover, we are able to prove that this greedy algorithm allows us to determine an upper bound in the case of three sources, which can be shown to be tight, i.e., there exists a network topology such that the cost gap is met with equality.
%
%
%
\subsection{Analysis of the gap between the coded case and the subset case}
%

\noindent We now present the problem formulation for ATOM-CODED-MIN-COST. We use the augmented graph $G^*$ in Figure \ref{fig:sour_stru}: 
%
%

minimize $\sum_{(i,j)\in E}f_{ij}z_{ij}+\sum_{i\in S}d_iz_{s^*i}$

subject to $0\leq x^{(t)}_{ij}\leq z_{ij}\leq c_{ij}^*,\forall (i,j)\in E^*, t\in T$
\begin{align}
\label{eq:chec}
&\sum_{\{j|(i,j)\in E^*\}}x_{ij}^{(t)}-\sum_{\{j|(j,i)\in E^*\}}x_{ji}^{(t)}=\sigma_i^{(t)},\forall i\in V^*,~t\in T\\
\label{eq:mincutc}
&x^{(t)}_{s^*i}\geq R_i^{(t)},\forall i\in S, t\in T
\end{align}
\begin{align}
\label{eq:swc}
&R^{(t)}\in \mathcal R_{\mathcal{SW}}, \forall t\in T\\
\label{eq:ele1c}
&H(X_i|X_{S\setminus\{i\}})\geq0,\forall i\in S\\
\label{eq:ele1c2}
&I(X_i;X_j|X_K)\geq0,\forall i\in S, j\in S, i\neq j, K\subseteq S\setminus\{i,j\}\\
\label{eq:marginalc}
&z_{s^*i}=H(X_i),\forall i\in S;~~~
H(X_1,X_2\cdots ,X_n)=\sum_{A\in\calA}\mu^*(A)
\end{align}
where $\sigma_i^{t}$ is defined in (\ref{eq:sigma}).
The formulation is the same as SUBSET-MIN-COST (Equation (\ref{eq:che})) except that we remove (\ref{eq:subset}), and add (\ref{eq:ele1c}) and (\ref{eq:ele1c2}), that are elemental inequalities, which guarantee that all Shannon type inequalities are satisfied \cite{first_info}. The constraints in (\ref{eq:ele1c}) and (\ref{eq:ele1c2}) can be represented in the form of atoms:
\begin{equation*}
\begin{split}
&H(X_i|X_{S\setminus\{i\}})=\mu^*(A),A\nsubseteq\wt{X}_{S\setminus\{i\}}\\
&I(X_i;X_j|X_K)=\sum_{A\in\calA:A\subset\wt{X}_i,A\subset\wt{X}_j,A\nsubseteq\wt{X}_K}\mu^*(A)
\end{split}
\end{equation*}
where $K\subseteq S\setminus\{i,j\}$.

Now we prove that ATOM-CODED-MIN-COST and CODED-MIN-COST have the same optimums. Let the optimum of ATOM-CODED-MIN-COST (CODED-MIN-COST) be $f_{opta}$ ($f_{optc}$). Denote $ConA=\{\text{the set of constraint of ATOM-CODED-MIN-COST}\}$ and $ConC=\{\text{the set of constraint of CODED-MIN-COST}\}$. First we note that the two LPs have the same objective functions, and $ConC\subset ConA$. Therefore, we should have $f_{opta}\geq f_{optc}$. Next we note that $\mu^*(A),A\in \calA$ are variables in $ConA\setminus ConC$ ((\ref{eq:mincutc})(\ref{eq:swc})(\ref{eq:ele1c})(\ref{eq:ele1c2})(\ref{eq:marginalc})). Let the optimal set of flows for CODED-MIN-COST be denoted as $x_{ij,c}^{(t)},z_{ij,c},t\in T, (i,j)\in E^*$. Now suppose that $f_{opta}>f_{optc}$. Note that this assignment is infeasible for ATOM-CODED-MIN-COST, since $f_{opta}>f_{optc}$. Next, since $ConC\subset ConA$, the constraints that cause infeasibility have to be in (\ref{eq:mincutc})-(\ref{eq:marginalc}). This implies that a feasible $\mu^*(A),A\in\calA$ cannot be found.

We claim that this is a contradiction. This is because if coding is allowed at the source, then there exists a deterministic algorithm \cite{Jaggi05} for the multicast network code assignment with a virtual source connected to all the source nodes that operates with the subgraph induced by $z_{ij,c},(i,j)\in E^*$. This algorithm guarantees the existence of random variables $X_1,\dots,X_n$ that correspond to the sources. This in turn implies the existence of atom measures that satisfy all information inequalities corresponding to the flow assignment $z_{ij,c},(i,j)\in E^*$. In the above LP, we have only enforce the elemental inequalities, therefore the existence for $\mu^*(A), A\in\calA$ is guaranteed.


Now, suppose that we know the optimal value of the above optimization problem, i.e., the flows $x_{ij,1}^{(t)}, z_{ij,1}^{(t)}, t\in T, (i,j)\in E^*$, the measure of the atoms $\mu^*(A)_{1},\forall A\in\calA$, and the corresponding conditional entropies $H^1(X_U|X_{S\setminus U}),\forall U\subseteq S$. If we can construct a feasible solution for SUBSET-MIN-COST such that the flows over $E^*$ are the same as $x_{ij,1}^{(t)} (\text{and~} z_{ij,1}^{(t)}), t\in T, (i,j)\in E$, then we can arrive at an upper bound for the gap. This is done below.

Let $\mu^*(A)$ denote the variables for the atom measures for the subset case. The gap LP is,

minimize
$$\sum_{A\in \calA}(\sum_{\{i \in S: A \subset \wt{X}_i\}} d_i)\mu^*(A)-\sum_{A\in \calA}(\sum_{\{i \in S: A \subset \wt{X}_i\}} d_i)\mu^*(A)_1$$

subject to
\begin{align}
\label{eq:remainflow}
\sum_{A:A\in \calA, A\nsubseteq\wt{X}_{S\setminus U}}\mu^*(A)&\leq H^1(X_U|X_{S\setminus U}),\forall U\subset S\\
\mu^*(A)&\geq0,\forall A\in\calA\nonumber\\
\sum_{A:A\in\calA}\mu^*(A)&=H(X_1,X_2,\cdots,X_n)\nonumber
\end{align}
where $H^1(X_U|X_{S\setminus U})=\sum_{A:A\in \calA,A\nsubseteq\wt{X}_{S\setminus U}}\mu^*(A)_1,\forall U\subset S$.
In the SUBSET-MIN-COST, we assign $x_{ij}^{(t)}=x_{ij,1}^{(t)}, (i,j)\in E^*$, $z_{ij}^{(t)} = z_{ij,1}^{(t)}, (i,j) \in E$ and $z_{s^*i}=\sum_{A:A\in \calA, A\subset\wt{X}_i}\mu^*(A), \forall i \in S$. To see that this is feasible, note that

\begin{equation*}
\begin{split}
&z_{s^*i}=\sum_{A:A\in \calA, A\subset\wt{X}_i}\mu^*(A)=H(X_i)\\
&=H(X_1,\cdots,X_n)-H(X_1,\cdots,X_{i-1},X_{i+1},\cdots,X_n|X_i)\\
&\stackrel{_{(a)}}{_{\geq}} H(X_1,\cdots,X_n)-H^1(X_1,\cdots,X_{i-1},X_{i+1},\cdots,X_n|X_i)\\
&=H^1(X_i)=z_{s^*i,1}\\
&\geq x_{s^*i,1}^{(t)}=x_{s^*i}^{(t)}.
\end{split}
\end{equation*}

This implies that constraint (\ref{eq:che}) is satisfied.
\vspace{-0.1in}
\begin{equation*}
\begin{split}
\sum_{i:i\in U}x_{s^*i}^{(t)}&=\sum_{i:i\in U}x_{s^*i,1}^{(t)}\geq H^1(X_U|X_{S\setminus U})\stackrel{_{(b)}}{\geq}H(X_U|X_{S\setminus U})
\end{split}
\end{equation*}
where $H(X_U|X_{S\setminus U})=\sum_{A:A\in \calA, A\nsubseteq\wt{X}_{S\setminus U}}\mu^*(A),\forall U\subset S$. Then constraints (\ref{eq:mincut}) and (\ref{eq:sw}) are satisfied.

Both $(a)$ and $(b)$ come from constraint (\ref{eq:remainflow}). The difference in the costs is only due to the different storage costs, since the flow costs are exactly the same. It is possible that the atom measures from ATOM-CODED-MIN-COST are not valid since they may not satisfy the non-Shannon inequalities. However, we claim that the solution of the Gap LP is still an upper bound of the difference between the coded and the subset case. This is because (a) we have constructed a feasible solution for SUBSET-MIN-COST starting with $\mu^*(A)_1,\forall A\in\calA$, and (b), as argued above, the optimal values of CODED-MIN-COST and ATOM-CODED-MIN-COST are the same. The difference between the costs in the coded case and the subset case are only due to the different storage costs, since the flows in both cases are the same. Therefore, the objective function of the gap LP is a valid upper bound on the gap. 


\subsection{Greedy Algorithm}
\label{sec:greedy}
We present a greedy algorithm for the gap LP that returns a feasible, near-optimal solution, and hence serves as an upper bound to the gap.
The main idea is to start by saturating atom values with the low costs, while still remaining feasible. For instance, suppose that source $1$ has the smallest cost. Then, the atom $\wt{X}_1 \cap_{k \in \calN_S \setminus \{1\}} \wt{X}_k^c$ has the least cost, and therefore we assign it the maximum value possible, i.e., $H^1(X_1 | X_{S \setminus \{1\}})$. Further assignments are made similarly in a greedy fashion. More precisely, we follow the steps given below.
\begin{enumerate}
\item Initialize $\mu^*(A) = 0, \forall A \in \calA$. Label all atoms as ``unassigned".
\item If all atoms have been assigned, STOP. Otherwise, let $A_{\min}$ denote the atom with the minimum cost that is still unassigned.
\begin{itemize}
\item Set $\mu^*(A_{\min}) \geq 0$ as large as possible so that the sum of the values of all assigned atoms does not violate any constraint in (\ref{eq:remainflow}).
\item Check to see whether $\sum_{A \in \calA} \mu^*(A) > H(X_1, X_2, \cdots, X_n)$. If YES, then reduce the value of $\mu^*(A_{\min})$, so that $\sum_{A \in \calA} \mu^*(A) = H(X_1, X_2, \cdots, X_n)$ and STOP. If NO, then label $A_{\min}$ as ``assigned".
\end{itemize}
\item Go to step 2.
\end{enumerate}
It is clear that this algorithm returns a feasible set of atom values, since we maintain feasibility at all times and enforce the sum of the atom values to be $H(X_1, X_2, \cdots, X_n)$.

The greedy algorithm, though suboptimal, does give the exact gap in many cases that we tested. Moreover, as discussed next, the greedy approach allows us to arrive at a closed form expression for the an upper bound on the gap in the case of three sources. However, it is not clear if there is a constant factor approximation for the greedy algorithm.
\subsection{Three sources case}
\label{sec:3sources}
The case of three sources is special because, (i) Shannon type inequalities suffice to describe the entropic region, i.e., non-Shannon type inequalities do not exist for three random variables. This implies that we can find three random variables using the atom measures found by the solution of ATOM-CODED-MIN-COST. (ii) Moreover, there is at most one atom measure, $\mu^*(\widetilde{X}_1\cap \widetilde{X}_2\cap \widetilde{X}_3)$ that can be negative. This makes the analysis easier since the greedy algorithm proposed above can be applied to obtain the required bound. Let $b=\mu^*(\widetilde{X}_1\cap \widetilde{X}_2\cap \widetilde{X}_3)$, $a_1=\mu^*(\widetilde{X}_1\cap \widetilde{X}_2^c\cap \widetilde{X}_3^c)$, $a_2=\mu^*(\widetilde{X}_2\cap \widetilde{X}_1^c\cap \widetilde{X}_3^c)$, $a_3=\mu^*(\widetilde{X}_3\cap \widetilde{X}_2^c\cap \widetilde{X}_1^c)$, $a_4=\mu^*(\widetilde{X}_1\cap \widetilde{X}_2\cap \widetilde{X}_3^c)$, $a_5=\mu^*(\widetilde{X}_1\cap \widetilde{X}_3\cap \widetilde{X}_2^c)$, and $a_6=\mu^*(\widetilde{X}_2\cap \widetilde{X}_3\cap \widetilde{X}_1^c)$. 

\begin{claim}
Consider random variables $X_1, X_2$ and $X_3$ with $H(X_1, X_2, X_3) = h$. Then, $b \geq -\frac{h}{2}$.
\end{claim}
\begin{proof}
The elemental inequalities are given by $a_i\geq0,i=1,\cdots,6$ (non-negativity of conditional entropy and conditional mutual information) and $a_i+b\geq0,i=4,5,6$ (non-negativity of mutual information). We also have $(\sum_{i=1,\cdots,6}a_i)+b=h$. Assume that $b<-\frac{h}{2}$. Then,
$$
a_i+b\geq0\Rightarrow a_i\geq-b>\frac{h}{2},i=4,5,6\Rightarrow a_4+a_5>h.
$$
\noindent Next,
\begin{equation*}
\begin{split}
h&=a_1+a_2+a_3+a_4+a_5+a_6+b\\
&\geq a_1+a_2+a_3+a_4+a_5>a_1+a_2+a_3+h.
\end{split}
\end{equation*}
\noindent This implies that $a_1+a_2+a_3<0$, which is a contradiction, since $a_i\geq0,i=1,\cdots,6$.
\end{proof}

Using this we can obtain the following lemma
\begin{lemma}
Suppose that we have three source nodes. Let the joint entropy of the original source be $h$ and let $f_{opt2}$ represent the optimal value of SUBSET-MIN-COST and $f_{opt1}$, the optimal value of CODED-MIN-COST. Let $b^*$ and $a_i^*$ be the optimal value of $b$ and $a_i$ in the coded case, respectively. If $b^*\geq 0$, the costs for the coded case and the subset case will be the same. If $b^*< 0$,  $f_{opt2}-f_{opt1}\leq (\min_{i \in S} (d_i))\times|b^*| \leq (\min_{i \in S} (d_i)) h/2 $.
\end{lemma}
\begin{proof}
When $b^*\geq 0$, the subset case atom values equal to the coded case atom values, then the two cases have the same costs. When $ b^* \leq 0$, without loss of generality, assume that $\min_{i \in S} (d_i) = d_1$. As in the greedy algorithm above, we construct a feasible solution for SUBSET-MIN-COST by keeping the flow values the same, but changing the atom values suitably. Let $a^2_i, i = 1, \dots, 6, b^2$ denote the atom values for the subset case. Consider the following assignment,
\begin{align*}
a^2_i = a^*_i,i=1, \dots, 5;~~~~~~
a^2_6 = a^*_6-|b^*|;~~~~~~
b^2=0.
\end{align*}

This is shown pictorially in Figure \ref{fig:t2}.
\begin{figure}[htbp]
\begin{center}
\includegraphics[width=90mm,clip=false, viewport=0 0 400 250 ]{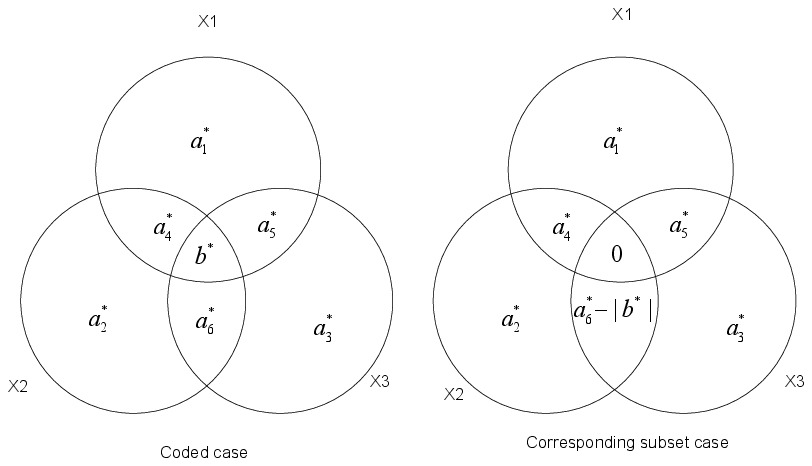}
\caption{The figure illustrates a transformation from the coded case to the subset case, when the first source has the minimal storage cost and $b^*<0$.} \label{fig:t2}\centering
\end{center}
\end{figure}

We can check constraint (\ref{eq:remainflow}) to see that the solution is feasible for the gap LP for three sources. It can also be verified that we can arrive at the above assignment by applying our greedy algorithm. Furthermore, on checking the KKT conditions of the gap LP, we conclude that the obtained solution is the optimal solution for the gap LP.
$x^{(t)}_{ij,1},(i,j)\in E^*$ are feasible for the subset problem. The flows do not change over transforming the coded case to the subset case. The only cost increased by transforming from the coded case to the subset case is $d_1\times |b^*|\leq (\min_{i \in S} (d_i)) h/2$.
\end{proof}

In the results section, we shall show an instance of a network where this upper bound is tight.
%
%

Finally we note that, when there are only two source nodes, there is no cost difference between the subset case and the coded case, since for two random variables, all atoms have to be nonnegative. We state this as a lemma below.
\begin{lemma}
Suppose that we have two source nodes. Let $f_{opt2}$ represent the optimal value of SUBSET-MIN-COST and $f_{opt1}$, the optimal value of CODED-MIN-COST. Then, $f_{opt2}=f_{opt1}$.
\end{lemma}
\section{Simulation results}
\label{sec:results}
\begin{figure}[t]
\begin{center}
\includegraphics[width=70mm,clip=false, viewport=0 0 350 210]{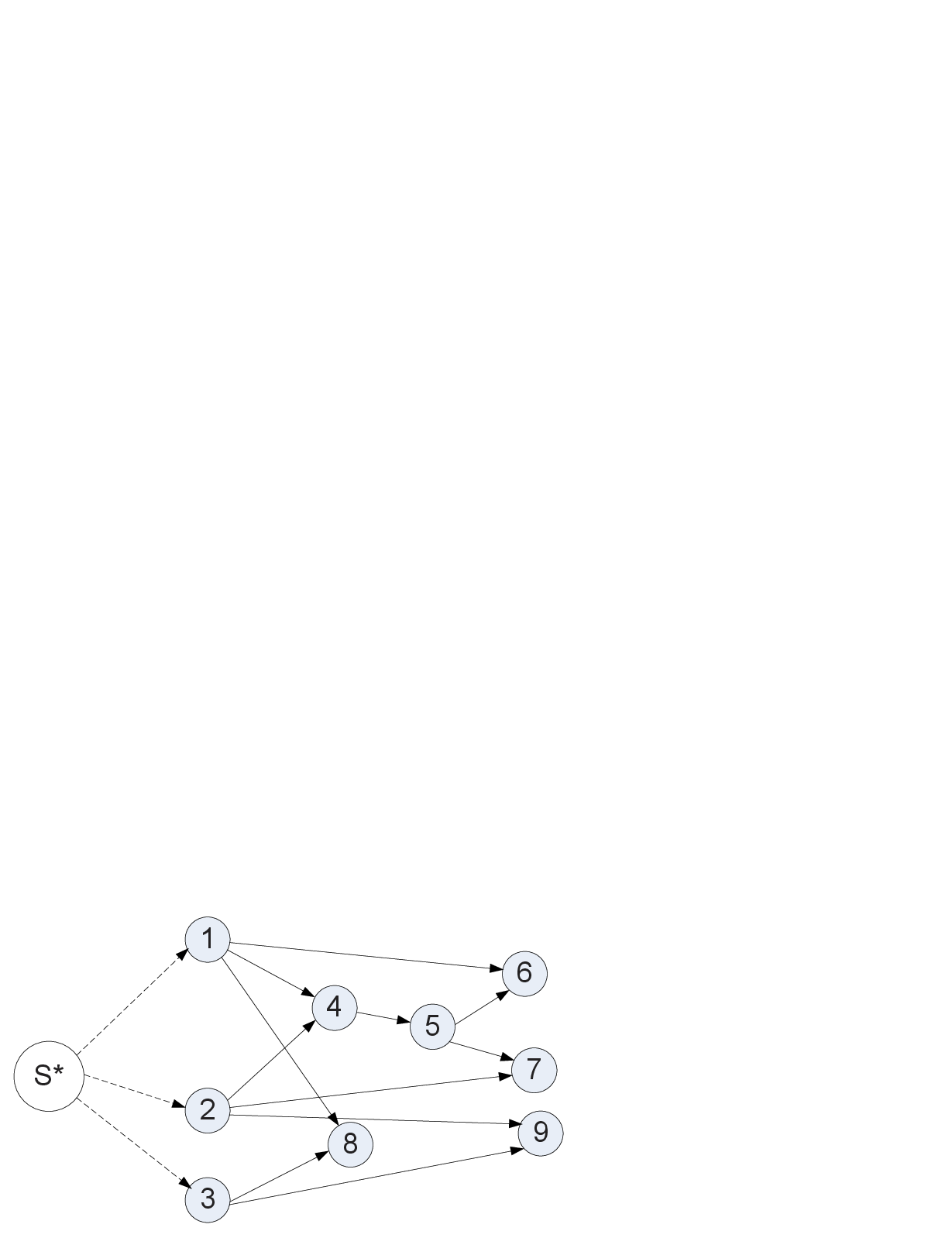}
\caption{Network with source nodes at 1, 2 and 3; terminals at 6, 7, 8 and 9. Append a virtual source $S^*$ connecting real sources.} \label{fig:eg}\centering
\end{center}
\end{figure}
In this section we present an example of a network with three sources where our upper bound derived in Section \ref{sec:3sources} is tight. We also present results of several experiments with randomly generated graphs. The primary motivation was to study whether the difference in cost between the subset sources case and the coded case occurs very frequently.

Consider the network in Figure \ref{fig:eg} with three sources nodes, 1, 2 and 3 and four terminal nodes, 6, 7, 8, and 9. The entropy of the original source = $H(X_1, X_2, X_3) = 2$ and all edges are unit-capacity. The costs are such that $f_{ij} = 1, \forall (i,j) \in E$ and $d_1 = d_2 = 2, d_3 = 1$.

The optimal cost in the subset sources case is $17$. The corresponding atom values are listed in the Table \ref{table:atom}. In this case we have 

\begin{table}
\scriptsize \caption{Atom values when subset constraints are enforced } \centering
{\begin{tabular}{c|c}
\hline \\
{\bf Atom} & $\mu^*(\cdot)$\\
\hline\\
{\bf $\wt{X}_1\cap\wt{X}_2^c\cap\wt{X}_3^c$} & 0\\
\hline\\
{\bf $\wt{X}_1^c\cap\wt{X}_2\cap\wt{X}_3^c$} &0\\
\hline\\
{\bf $\wt{X}_1\cap\wt{X}_2\cap\wt{X}_3^c$} &0.5809\\
\hline\\
{\bf $\wt{X}_1^c\cap\wt{X}_2^c\cap\wt{X}_3$} &0\\
\hline\\
{\bf $\wt{X}_1\cap\wt{X}_2^c\cap\wt{X}_3$} &0.6367\\
\hline\\
{\bf $\wt{X}_1^c\cap\wt{X}_2\cap\wt{X}_3$} &0.7824\\
\hline\\
{\bf $\wt{X}_1\cap\wt{X}_2\cap\wt{X}_3$}&0\\
\hline
\end{tabular}}
\label{table:atom} 
\end{table} \normalsize

In the coded sources case, the optimal value is 16, with $H(X_1)=H(X_2)=H(X_3)=1$. Note that in this case the gap between the optimal values is precisely = $\frac{2}{2} \times 1 = 1$, i.e., the upper bound derived in the previous section is met with equality.

%

We generated several directed graphs at random with $|V|=87$,
$|E|=322$. The linear cost of each edge was fixed to an integer in
$\{1,2,3,4,5,6,29,31\}$. We ran 5000 experiments with fixed
parameters $(|S|,|T|,h)$, where $|S|$ - number of source nodes,
$|T|$ - number of terminal nodes, and $h$ - entropy of the
original source. The locations of the source and terminal nodes
were chosen randomly. The capacity of each edge was chosen at
random from the set $\{1,2,3,4,5\}$. In many cases it turned out
that the network did not have enough capacity to support recovery
of the data at the terminals. These instances were discarded.

The results are shown in Table \ref{table:results}.
The ``Equal" row corresponds to the number of instances when both the coded and subset cases have the same cost, and ``Non-equal" corresponds to the number of instances where the coded case has a lower cost. We have found that in most cases, the two cases have the exact same cost.
%
We also computed the gap LP and the greedy algorithm 
to evaluate the cost gap. Note that the gap LP is only an upper bound since it is derived assuming that the flow patterns do not change between the two cases. When $(|S|,|T|,h)=(4,3,4)$, among 5000 experiments, 3269 instances could support both cases. Out of these, there were 481 instances where the upper bound determined by the gap LP was not tight. In addition, there were 33 instances where the greedy algorithm failed to solve the gap LP exactly.
\begin{table}
\scriptsize \caption{Comparisons of two schemes in 5000 random directed graphs} \centering
\resizebox{9cm}{!}
{\Large\begin{tabular}{c| c c c c c c}
\hline \\
{$(|S|,|T|,h)$} & {\bf $(3,3,3)$} & {\bf $(4,4,4)$} & {\bf $(5,5,5)$}& {\bf $(4,5,5)$}& {\bf $(5,4,5)$}& {\bf $(4,4,5)$}\\
\hline\hline\\
$Equal$&3893&2855&1609&1577&2025&1954\\
\hline\\
$Non-equal$&1&3&10&9&6&8\\
\hline
\end{tabular}}
\label{table:results} 
\end{table} \normalsize
\section{Conclusions and Future work}
\label{sec:conclu}
%

In this work, we considered network coding based content
distribution, under the assumption that the content can be
considered as a collection of independent equal entropy sources.
e.g., a large file that can be subdivided into small pieces. Given
a network with a specified set of source nodes, we examined two
cases. In the subset sources case, the source nodes are
constrained to only contain subsets of the pieces of the content,
whereas in the coded sources case, the source nodes can contain
arbitrary functions of the pieces. The cost of a solution is
defined as the sum of the storage cost and the cost of the flows
required to support the multicast. We provided succinct
formulations of the corresponding optimization problems by using
the properties of information measures. In particular, we showed
that when there are two source nodes, there is no loss in
considering subset sources. For three source nodes, we derived a
tight upper bound on the cost gap between the two cases. A greedy
algorithm for estimating the cost gap for a given instance was
provided. Finally, we also provided algorithms for determining the
content of the source nodes. Our results indicate that when the
number of source nodes is small, in many cases constraining the
source nodes to only contain subsets of the content does not incur
a loss.

In our work, we have used linear objective functions. However,
this is not necessary. We could also have used convex functions.
That would simply not have allowed a LP formulation and the gap
bound would be different. In our work, we have assumed that the
locations of the source node are known. It would be interesting to
consider, whether one can extend this work to identify the optimal
locations of the source nodes, e.g., if an ISP wants to establish
mirror sites, what their geographical locations should be. The gap
between subset and coded sources shown here is for three sources.
It would be interesting to see how it grows with the number of
sources. We conjecture that the gap can be quite large when the
number of source nodes is high. We have investigated the difference of the coded and subset case under a network with arbitrary topology. Examining this issue when the network has structural constraints (such as bounded treewidth \cite{bounded_treewidth}) could be another avenue for future work.

Beyond gaps, there may be advantages to coding when we have
multi-tiered distributed storage, such as in the case in current
large CDNs. In that case, the subset approach would require extra
constraints in the middle tiers that may be difficult to keep
track of. The coded storage approach gracefully extends to a
multi-tiered architecture.

\section{Acknowledgements}

The authors would like to thank of the anonymous reviewers whose comments greatly improved the quality and presentation of the paper.

\appendix
\subsection{Proof of Theorem \ref{th:sb}.}
(1)
Independent random variables $W_A, A \in \calA$, such that $H(W_A)=\alpha(A)$ can be constructed \cite{first_info}. Then we can set $X_i=(W_A:A\in\calA, A\subset\wt{X}_i)$. It only remains to check the consistency of the measures. For this, we have, for all $V \subseteq \calN_S$,
\begin{equation}
H(X_V) = \sum_{A \in \calA:A \subset \wt{X}_V} H(W_A),
\end{equation}
using the independence of the $W_A$'s. On the other hand we know that
\begin{equation}
H(X_V) = \mu^*(\wt{X}_V) = \sum_{A \in \calA: A \subset \wt{X}_V} \mu^*(A).
\end{equation}
Equating these two we have, for all $V \subseteq \calN_S$,
\begin{equation}
\sum_{A \in \calA:A \subset \wt{X}_V} H(W_A) = \sum_{A \in \calA: A \subset \wt{X}_V} \mu^*(A).
\end{equation}
Now, one possible solution to this is that $\mu^*(A) = H(W_A), \forall A \in \calA$. By the uniqueness of $\mu^*$ \cite{first_info}, we know that this is the only solution.

(2) 
We shall prove all the measures are nonnegative by induction. Without loss of generality, 
we can order $\wt{X}_i$'s in an arbitrary way, we analyze the measure $\mu^*(\widetilde{X}_1\cap\cdots\cap\widetilde{X}_l\cap_{k:k\in K}\widetilde{X}_k^c)$
where $K \subseteq \calN_S \setminus \{1,2,\cdots, l\}$, $l\leq n$.

When $l=1$, the measure corresponds to conditional entropy, $\forall K\subseteq \mathcal N_S\setminus\{1\}$
$$\mu^*(\widetilde{X}_1\cap_{k:k\in K}\widetilde{X}_k^c)=H(X_1|X_K)\geq0.$$

When $l=2$, we have, $\forall K\subseteq \mathcal N_S\setminus\{1,2\}$
\begin{equation*}
\begin{split}
&\mu^*(\widetilde{X}_1\cap\widetilde{X}_{2}\cap_{k:k\in K}\widetilde{X}_k^c)=I(X_1;X_2|X_K)\\
&=H(X_1,X_K)+H(X_2,X_K)-H(X_K)-H(X_1,X_2,X_K)\\
&=\sum_{i\in{V_1\cap V_2\cap_{k:k\in K} V_k^c}}H(Z_i)\geq0.
\end{split}
\end{equation*}

Assume for $l=j$, $\forall K\subseteq \mathcal N_S\setminus\{1,2,\cdots,j\}$, the following statement holds,
\begin{equation}
\label{eq:assum}
\mu^*(\widetilde{X}_1\cap\cdots\cap\widetilde{X}_j\cap_{k:k\in K}\widetilde{X}_k^c)
=\sum_{i\in{V_1\cap\cdots\cap V_j\cap_{k:k\in K} V_k^c}}H(Z_i).
\end{equation}

When $l=j+1$, $\forall K\subseteq \mathcal N_S\setminus\{1,2,\cdots,j+1\}$, we shall have
\begin{equation*}
\begin{split}
&\mu^*(\widetilde{X}_1\cap\cdots\cap\widetilde{X}_{j+1}\cap_{k:k\in K}\widetilde{X}_k^c)\\
&=\mu^*(\widetilde{X}_1\cap\cdots\cap\widetilde{X}_{j}\cap_{k:k\in K}\widetilde{X}_k^c)\\
&~~~~~~-\mu^*(\widetilde{X}_1\cap\cdots\cap\widetilde{X}_j\cap\widetilde{X}_{j+1}^c\cap_{k:k\in K}\widetilde{X}_k^c)\\
& \stackrel{_{(a)}}{=}\sum_{i\in{V_1\cap\cdots\cap V_j\cap_{k:k\in K} V_k^c}}H(Z_i)
-\sum_{i\in{V_1\cap\cdots\cap V_j\cap V_{j+1}^c\cap_{k:k\in K} V_k^c}}H(Z_i)\\
& \stackrel{_{(b)}}{=}\sum_{i\in{V_1\cap\cdots\cap V_{j+1}\cap_{k:k\in K} V_k^c}}H(Z_i)\geq0.
\end{split}
\end{equation*}
The equation $(a)$ is due to the assumption (\ref{eq:assum}). The equation $(b)$ is due to the independence of $Z_i$'s, $i\in\{1, \dots, m\}$.
Therefore, we have shown that $j\leq n$, $\forall K\subseteq \mathcal N_S\setminus\{1,2,\cdots,j\}$,
\vspace{-0.1in}
\begin{equation*}
\begin{split}
\mu^*(\widetilde{X}_1\cap&\cdots\cap\widetilde{X}_j\cap_{k:k\in K}\widetilde{X}_k^c)
=\sum_{i\in{V_1\cap\cdots\cap V_j\cap_{k:k\in K} V_k^c}}H(Z_i)\geq0.
\end{split}
\end{equation*}
In a similar manner it is easy to see that all atom measures are non-negative.
%
%
\vspace{-0.2in}
\bibliographystyle{IEEEtran}

\begin{thebibliography}{10}
\providecommand{\url}[1]{#1}
\csname url@samestyle\endcsname
\providecommand{\newblock}{\relax}
\providecommand{\bibinfo}[2]{#2}
\providecommand{\BIBentrySTDinterwordspacing}{\spaceskip=0pt\relax}
\providecommand{\BIBentryALTinterwordstretchfactor}{4}
\providecommand{\BIBentryALTinterwordspacing}{\spaceskip=\fontdimen2\font plus
\BIBentryALTinterwordstretchfactor\fontdimen3\font minus
  \fontdimen4\font\relax}
\providecommand{\BIBforeignlanguage}[2]{{%
\expandafter\ifx\csname l@#1\endcsname\relax
\typeout{** WARNING: IEEEtran.bst: No hyphenation pattern has been}%
\typeout{** loaded for the language `#1'. Using the pattern for}%
\typeout{** the default language instead.}%
\else
\language=\csname l@#1\endcsname
\fi
#2}}
\providecommand{\BIBdecl}{\relax}
\BIBdecl

\bibitem{kirpal00}
A.~Kirpal, P.~Rodriguez, and E.~W. Biersack, ``{Parallel-Access for Mirror
  Sites in the Internet},'' in \emph{INFOCOM 2000, Nineteenth Annual Joint
  Conference of the IEEE Computer and Communications Societies}, vol.~2, 2000,
  pp. 864--873.

\bibitem{p2pperfomance}
D.~Qiu and R.~Srikant, ``{Modeling and performance analysis of BitTorrent-like
  peer-to-peer networks},'' in \emph{Proceedings of the 2004 conference on
  Applications, technologies, architectures and protocols for computer
  communications}, 2004, pp. 367--378.

\bibitem{filestorage}
A.~Jiang and J.~Bruck, ``{Network File Storage with Graceful Performance
  Degradation},'' \emph{ACM Transactions on Storage}, vol. 1, no. 2, pp.
  171--189, 2005.

\bibitem{zhanghui03}
K.~Sripanidkulchai, B.~Maggs, and H.~Zhang, ``{Efficient Content Location Using
  Interest-Based Locality in Peer-to-Peer systems},'' in \emph{INFOCOM 2003.
  22rd Annual Joint Conference of the IEEE Computer and Communications
  Societies}, vol.~3, 2003, pp. 2166--2176.

\bibitem{CDN1}
Y.~Chen, L.~Qiu, W.~Chen, L.~Nguyen, and R.~H. Katz, ``{Efficient and adaptive
  Web replication using content clustering},'' in \emph{IEEE Journal on
  Selected Areas in Communications}, vol.~21, Aug. 2003, pp. 979--994.

\bibitem{CDN2}
K.~Hosanagar, R.~Krishnan, M.~Smith, and J.~Chuang, ``{Optimal pricing of
  content delivery network services},'' in \emph{Proceedings of 37th Annual
  Hawaii International Conference on System Sciences}, Jan. 2004, p.~10.

\bibitem{CDN3}
K.~L. Johnson, J.~F. Carr, M.~S. Day, and M.~F. Kaashoek, ``{The measured
  performance of content distribution networks},'' in \emph{Computer
  Communications}, vol. 24, No. 2, 2001, pp. 202--206.

\bibitem{akamai}
D.~Karger, E.~Lehman, T.~Leighton, M.~Levine, D.~Lewin, and R.~Panigrahy,
  ``{Consistent Hashing and Random Trees: Distributed Caching Protocols for
  Relieving Hot Spots on the World Wide Web},'' in \emph{ACM STOC}, 1997, pp.
  834--838.

\bibitem{ratnasamy01}
S.~Ratnasamy, M.~Handley, R.~Karp, and S.~Shenker, ``{A Scalable
  Content-Addressable Network},'' in \emph{ACM SIGCOMM}, 2001, pp. 161--172.

\bibitem{contentdist}
C.~Gkantsidis and P.~Rodriguez, ``{Network Coding for Large Scale Content
  Distribution},'' in \emph{INFOCOM 2005. 24th Annual Joint Conference of the
  IEEE Computer and Communications Societies}, vol.~4, 2005, pp. 2235--2245.

\bibitem{lunmincost}
D.~S. Lun, N.~Ratnakar, M.~M\'{e}dard, R.~Koetter, D.~R. Karger, T.~Ho,
  E.~Ahmed, and F.~Zhao, ``{Minimum-Cost Multicast over Coded Packet
  Networks},'' \emph{IEEE Trans. on Info. Th.}, vol.~52, pp. 2608--2623, June
  2006.

\bibitem{jain07}
K.~Jain, L.~Lov\'{a}sz, and P.~A. Chou, ``{Building scalable and robust
  peer-to-peer overlay networks for broadcasting using network coding},''
  \emph{Distributed Computing}, vol. 19(4), pp. 301--311, 2007.

\bibitem{zhao05}
J.~Zhao, F.~Yang, Q.~Zhang, Z.~Zhang, and F.~Zhang, ``{LION: layered overlay
  multicast with network coding},'' in \emph{IEEE Transactions on Multimedia},
  vol. 8, No. 5, Oct. 2006, pp. 1021--1032.

\bibitem{wuyunnanp2p}
Y.~Wu, Y.~C. Hu, J.~Li, and P.~A. Chou, ``{The Delay Region for P2P File
  Transfer},'' in \emph{IEEE Intl. Symposium on Info. Th.}, 2009, pp. 834--838.

\bibitem{Ho09delay}
T.~K. Dikaliotis, A.~G. Dimakis, T.~Ho, and M.~Effros, ``{On the Delay of
  Network Coding over Line Networks},'' in \emph{IEEE Intl. Symposium on Info.
  Th.}, 2009, pp. 1408--1412.

\bibitem{ezovski09}
G.~M. Ezovski, A.~Tang, and L.~L.~H. Andrew, ``{Minimizing Average Finish Time
  in P2P Networks},'' in \emph{INFOCOM}, 2009, pp. 594--602.

\bibitem{adisepDSC}
A.~Ramamoorthy, K.~Jain, P.~A. Chou, and M.~Effros, ``{Separating Distributed
  Source Coding from Network Coding},'' \emph{IEEE Trans. on Info. Th.},
  vol.~52, pp. 2785--2795, June 2006.

\bibitem{liR10_chapter}
S.~Li and A.~Ramamoorthy, ``Network distributed source coding,''
  \emph{Theoretical Aspects of Distributed Computing in Sensor Networks}, 2010.

\bibitem{netcod07}
A.~Lee, M.~M\'{e}dard, K.~Z. Haigh, S.~Gowan, and P.~Rubel, ``{Minimum-Cost
  Subgraphs for Joint Distributed Source and Network Coding},'' in \emph{the
  Third Workshop on Network Coding, Theory, and Applications}, Jan. 2007.

\bibitem{ramamoorthy09}
\BIBentryALTinterwordspacing
A.~Ramamoorthy, ``{Minimum cost distributed source coding over a network},'' in
  \emph{IEEE Trans. on Information Theory (to appear)}. [Online]. Available:
  \url{http://arxiv.org/abs/0704.2808}
\BIBentrySTDinterwordspacing

\bibitem{first_info}
R.~Yeung, \emph{Information Theory and Network Coding}.\hskip 1em plus 0.5em
  minus 0.4em\relax Springer, 2008.

\bibitem{baochunli08}
C.~Feng and B.~Li, ``{On large-scale peer-to-peer streaming systems with
  network coding},'' in \emph{ACM international conference on Multimedia,
  Vancouver, British Columbia, Canada}, 2008, pp. 269--278.

\bibitem{netcod2005_accek}
S.~Aceda\'{n}ski, S.~Deb, M.~M\'{e}dard, and R.~Koetter, ``{How Good is Random
  Linear Coding Based Distributed Networked Storage},'' in \emph{NetCod}, Apr.
  2005.

\bibitem{ramamoorthyRS09}
A.~Ramamoorthy, V.~P. Roychowdhury, and S.~K. Singh, ``{Selfish Distributed
  Compression over Networks},'' in \emph{Proc. of the $28^{th}$ IEEE Intl.
  Conf. on Computer Communications (INFOCOM) Mini-Conf.}, 2009.

\bibitem{jointopt}
A.~Jiang, ``{Network Coding for Joint Storage and Transmission with Minimum
  Cost},'' in \emph{IEEE Intl. Symposium on Info. Th.}, 2006, pp. 1359--1363.

\bibitem{NCDSS}
A.~G. Dimakis, P.~B. Godfrey, Y.~Wu, M.~Wainwright, and K.~Ramchandran,
  ``{Network Coding for Distributed Storage Systems},'' \emph{To appear on IEEE
  Transactions on Information Theory}.

\bibitem{minimal_nc}
K.~Bhattad, N.~Ratnakar, R.~Koetter, and K.~Narayanan, ``{Minimal Network
  Coding for Multicast},'' in \emph{IEEE Intl. Symposium on Info. Th.}, 2005,
  pp. 1730--1734.

\bibitem{NormalizedEV}
B.~Hassibi and S.~Shadbakht, ``{Normalized Entropy Vectors, Network Information
  Theory and Convex Optimization},'' in \emph{IEEE Information Theory Workshop
  for Information Theory for Wireless Networks, Solstrand, Norway}, July, 2007,
  pp. 1--6.

\bibitem{Tracy06}
T.~Ho, R.~Koetter, M.~M\'{e}dard, M.~Effros, J.~Shi, and D.~Karger, ``{A Random
  Linear Network Coding Approach to Multicast},'' \emph{IEEE Trans. on Info.
  Th.}, vol. 28, no. 4, pp. 585--592, 1982.

\bibitem{kleinburgT}
J.~Kleinberg and \'{E}va Tardos, \emph{Algorithm Design}.\hskip 1em plus 0.5em
  minus 0.4em\relax Person/Addison-Wesley, 2006.

\bibitem{Jaggi05}
S.~Jaggi, P.~Sanders, P.~A. Chou, M.~Effros, S.~Egner, K.~Jain, and L.~M. G.~M.
  Tolhuizen, ``{Polynomial Time Algorithms for Multicast Network Code
  Construction},'' \emph{IEEE Trans. on Info. Th.}, vol. 51, no. 6, pp.
  1973--1982, 2005.

\bibitem{infinite_matus}
F.~Mat\'{u}\v{s}, ``{Infinitely Many Information Inequalities},'' in \emph{IEEE
  Intl. Symposium on Info. Th.}, 2007, pp. 41--44.

\bibitem{bounded_treewidth}
H.~L. Bodlaender, ``{A partial k-arboretum of graphs with bounded treewidth},''
  \emph{Theoretical Computer Science}, vol. 209.1-2, pp. 1--45, Dec. 1998.

\end{thebibliography}

\begin{biography}
[{\includegraphics[width=1in,
height=1.25in,clip,keepaspectratio]{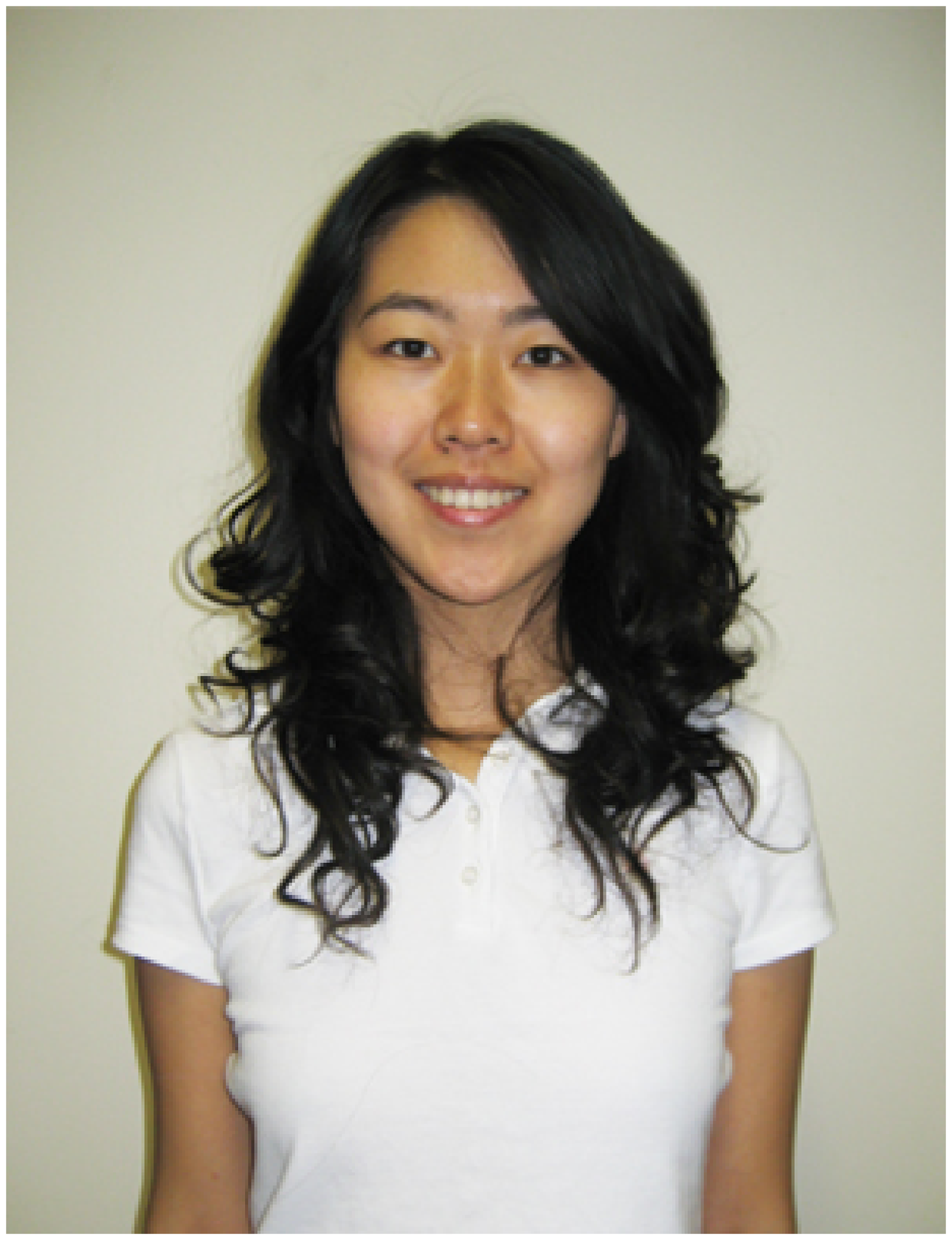}}]{Shurui Huang}Shurui Huang
received her B Eng degree in Automation from Tsinghua University, Beijing,
China in 2006. She worked on image processing in Broadband Network and Digital
Media Lab at Tsinghua University during 2005.
Since Fall 2006, she has been a Ph.D student in Department of
Electrical and Computer Engineering, Iowa State University. Her research
interests include channel coding, network coding theory and application.
\end{biography}

\begin{biography}[{\includegraphics[width=1in,
height=1.25in,clip,keepaspectratio]{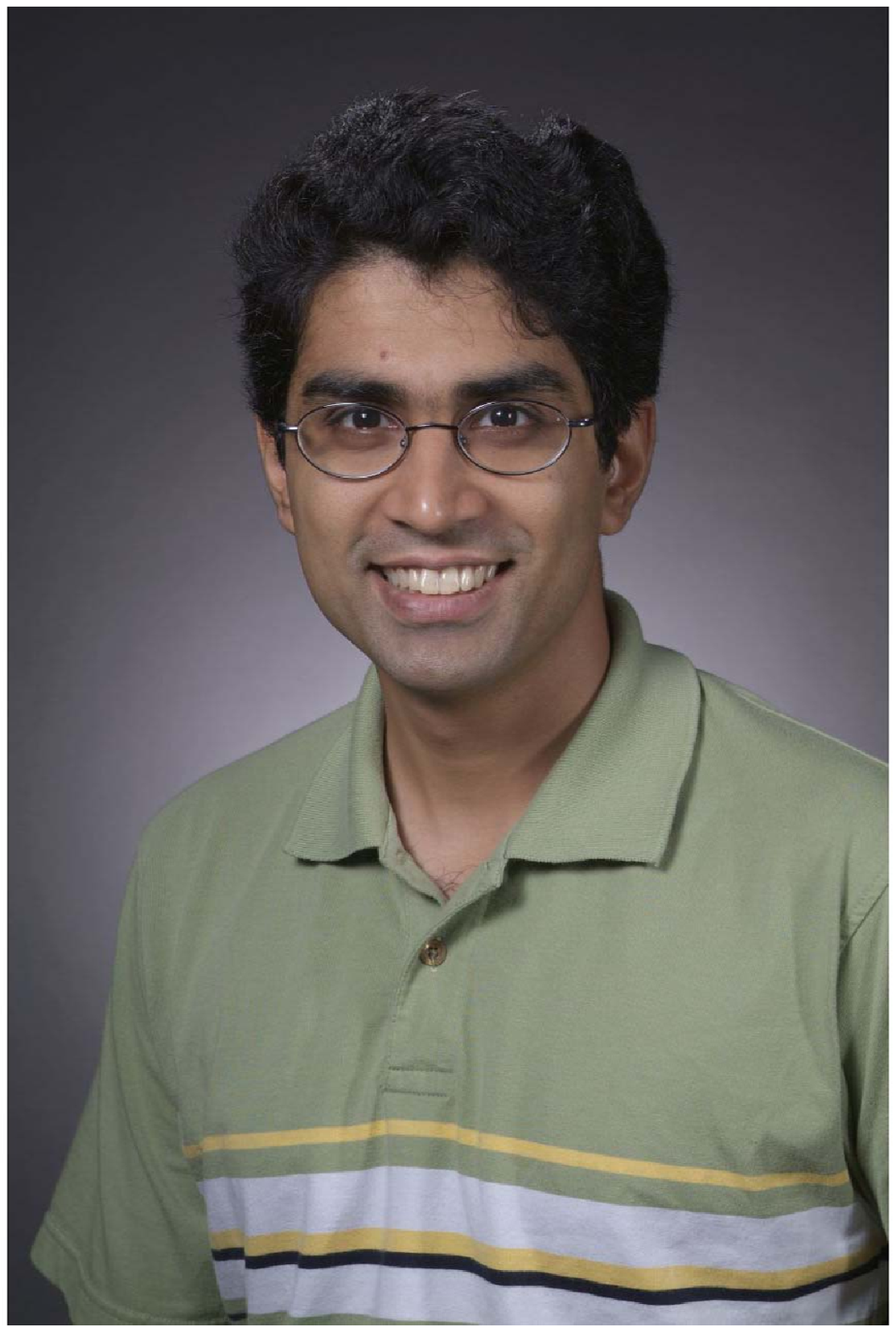}}]{Aditya
Ramamoorthy} Aditya Ramamoorthy received his B. Tech degree in
Electrical Engineering from the Indian Institute of Technology,
Delhi in 1999 and the M.S. and Ph.D. degrees from the University
of California, Los Angeles (UCLA) in 2002 and 2005 respectively.
He was a systems engineer at Biomorphic VLSI Inc. till 2001. From
2005 to 2006 he was with the data storage signal processing group
at Marvell Semiconductor Inc. Since Fall 2006 he has been an
assistant professor in the ECE department at Iowa State
University. His research interests are in the areas of network
information theory, channel coding, and signal processing for
storage devices and its application to nanotechnology.
\end{biography}

\begin{biography}[{\includegraphics[width=1in,
height=1.25in,clip,keepaspectratio]{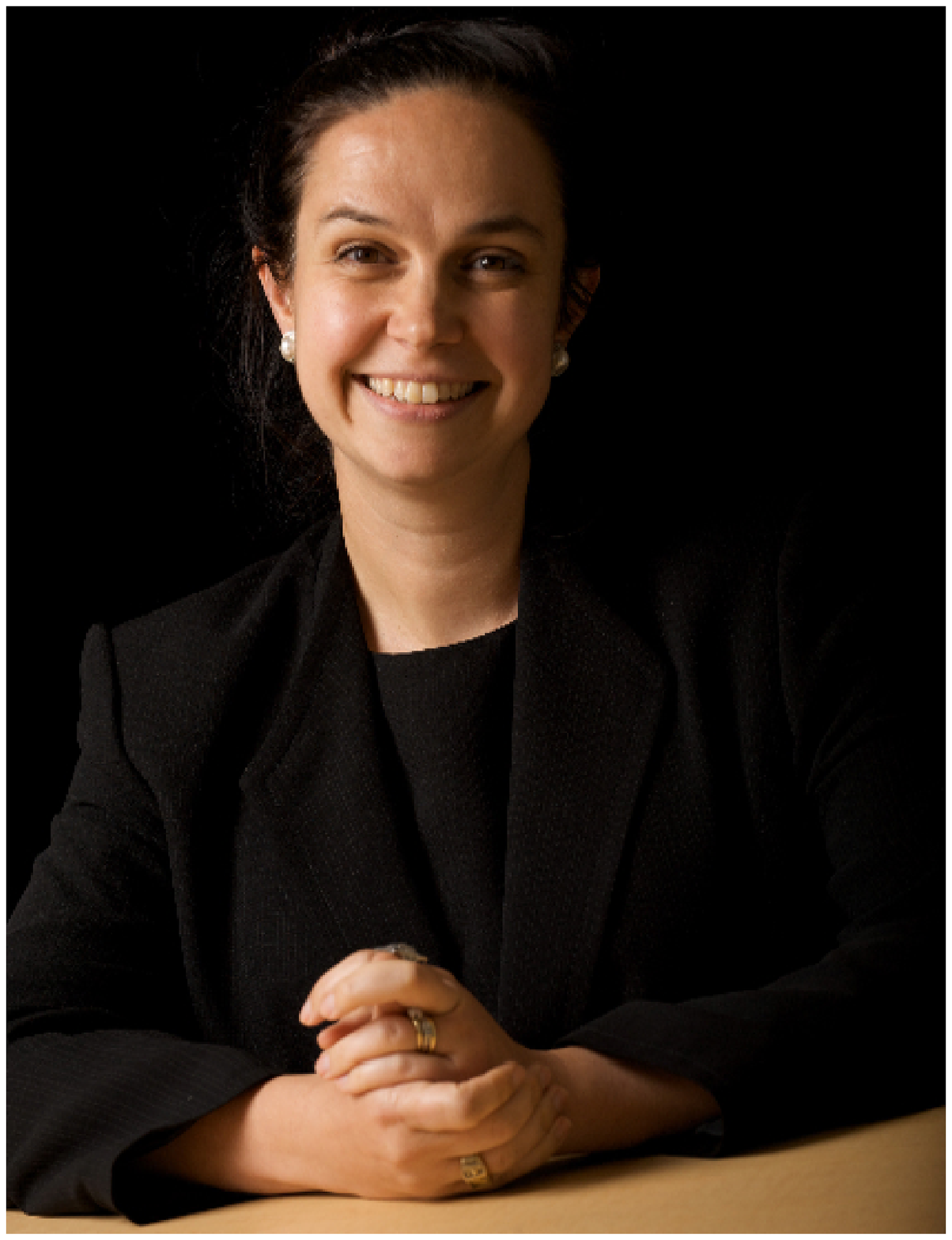}}]{Muriel M\'{e}dard}
Muriel M\'{e}dard is a Professor in the Electrical Engineering and Computer Science at MIT. She was previously an Assistant Professor in the Electrical and Computer Engineering Department and a member of the Coordinated Science Laboratory at the University of Illinois Urbana-Champaign. From 1995 to 1998, she was a Staff Member at MIT Lincoln Laboratory in the Optical Communications and the Advanced Networking Groups. Professor M\'{e}dard received B.S. degrees in EECS and in Mathematics in 1989, a B.S. degree in Humanities in 1990, a M.S. degree in EE 1991, and a Sc D. degree in EE in 1995, all from the Massachusetts Institute of Technology (MIT), Cambridge. She has served as an Associate Editor for the Optical Communications and Networking Series of the IEEE Journal on Selected Areas in Communications, as an Associate Editor in Communications for the IEEE Transactions on Information Theory and as an Associate Editor for the OSA Journal of Optical Networking. She has served as a Guest Editor for the IEEE Journal of Lightwave Technology, the Joint special issue of the IEEE Transactions on Information Theory and the IEEE/ACM Transactions on Networking on Networking and Information Theory and the IEEE Transactions on Information Forensic and Security: Special Issue on Statistical Methods for Network Security and Forensics. She serves as an associate editor for the IEEE/OSA Journal of Lightwave Technology. She is a member of the Board of Governors of the IEEE Information Theory Society.

Professor M\'{e}dard's research interests are in the areas of network coding and reliable communications, particularly for optical and wireless networks. She was awarded the 2009 Communication Society and Information Theory Society Joint Paper Award for the paper: Tracey Ho , Muriel M\'{e}dard, Rolf Kotter, David Karger, Michelle Effros Jun Shi, Ben Leong,  "A Random Linear Network Coding Approach to Multicast", IEEE Transactions on Information Theory, vol. 52, no. 10, pp. 4413-4430, October 2006. She was awarded the 2009 William R. Bennett Prize in the Field of Communications Networking for the paper: Sachin Katti , Hariharan Rahul, Wenjun Hu, Dina Katabi, Muriel M\'{e}dard, Jon Crowcroft, "XORs in the Air: Practical Wireless Network Coding", IEEE/ACM Transactions on Networking, Volume 16, Issue 3, June 2008, pp. 497 - 510. She was awarded the IEEE Leon K. Kirchmayer Prize Paper Award 2002 for her paper, "The Effect Upon Channel Capacity in Wireless Communications of Perfect and Imperfect Knowledge of the Channel," IEEE Transactions on Information Theory, Volume 46 Issue 3, May 2000, Pages: 935-946. She was co- awarded the Best Paper Award for G. Weichenberg, V. Chan, M. M\'{e}dard,"Reliable Architectures for Networks Under Stress", Fourth International Workshop on the Design of Reliable Communication Networks (DRCN 2003), October 2003, Banff, Alberta, Canada. She received a NSF Career Award in 2001 and was co-winner 2004 Harold E. Edgerton Faculty Achievement Award, established in 1982 to honor junior faculty members "for distinction in research, teaching and service to the MIT community."
\end{biography}

\end{document}